%% file: symbolic.tex
\documentclass[acmthm]{acmart}

\AtBeginDocument{%
  \providecommand\BibTeX{{%
    \normalfont B\kern-0.5em{\scshape i\kern-0.25em b}\kern-0.8em\TeX}}}

\setcopyright{acmcopyright}
\copyrightyear{2018}
\acmYear{2018}
\acmDOI{10.1145/1122445.1122456}

\acmJournal{TACO}
\acmVolume{xx}
\acmNumber{n}
\acmArticle{iii}
\acmMonth{3}

\usepackage{tikz}
\usetikzlibrary{positioning,calc,shapes}
\usetikzlibrary{shapes.geometric}
\usetikzlibrary{decorations, decorations.markings}
\usetikzlibrary{decorations.pathreplacing}
\usetikzlibrary{arrows.meta}
\usetikzlibrary{backgrounds}
\usetikzlibrary{patterns}

\usepackage{algorithm}
\usepackage[noend]{algpseudocode}
\usepackage{subcaption}

\usepackage{mathtools}
\usepackage{bm}
\usepackage{amsthm}
\usepackage{thmtools}
\DeclareMathOperator{\diag}{diag}
\declaretheoremstyle[
  headfont=\normalfont\bfseries\small,
  notefont=\mdseries\small,
  bodyfont=\normalfont\small]{example}
\declaretheorem[numbered=no,style=remark,qed=$\bigtriangleup$,name=Running example]{example}

\declaretheorem[numbered=yes,style=acmdefinition]{definition}

\let\vec\bm

\usepackage[inline]{enumitem}
\usepackage{scalerel}

\usepackage{booktabs}
\usepackage[keeplastbox]{flushend} % keeplastbox so last reference is correctly indented

\usepackage{forest}
\usepackage{textcomp}

\input{head/commands}

\input{head/tikz}
\input{head/acronyms}

\begin{document}

\title[Symbolic Loop Compilation for TCPAs]{Symbolic Loop Compilation\\for Tightly Coupled Processor Arrays}

\author{Michael Witterauf}
\email{michael.witterauf@fau.de}
\affiliation{%
  \institution{Friedrich-Alexander-Universität Erlangen-Nürnberg}
}
\author{Dominik Walter}
\email{dominik.l.walter@fau.de}
\affiliation{%
  \institution{Friedrich-Alexander-Universität Erlangen-Nürnberg}
}
\author{Frank Hannig}
\email{frank.hannig@fau.de}
\affiliation{%
  \institution{Friedrich-Alexander-Universität Erlangen-Nürnberg}
}
\author{Jürgen Teich}
\email{juergen.teich@fau.de}
\affiliation{%
  \institution{Friedrich-Alexander-Universität Erlangen-Nürnberg}
}

%%
%% By default, the full list of authors will be used in the page
%% headers. Often, this list is too long, and will overlap
%% other information printed in the page headers. This command allows
%% the author to define a more concise list
%% of authors' names for this purpose.
%\renewcommand{\shortauthors}{Trovato and Tobin, et al.}

%%
%% The abstract is a short summary of the work to be presented in the
%% article.
\begin{abstract}
\input{sections/abstract}
\end{abstract}

%%
%% The code below is generated by the tool at http://dl.acm.org/ccs.cfm.
%% Please copy and paste the code instead of the example below.
%%



\maketitle

\input{sections/introduction}
\input{sections/related-work}
\input{sections/tcpas}
\input{sections/prerequisites}
\input{sections/compile-time}
\input{sections/runtime}
\input{sections/discussion}
\input{sections/conclusion}

\bibliographystyle{ACM-Reference-Format}
\bibliography{literature}

\end{document}

%% file: head/commands.tex
\newcommand*{\keyword}[1]{\textit{#1}}
\newcommand*{\T}{\mathrm{T}}
\newcommand*{\annotate}[2]{#1[#2]}

\DeclareMathOperator{\attrs}{attr}

\DeclareMathOperator{\children}{children}
\newcommand*{\sinkport}[2]{\ensuremath{\mathit{#1}_{#2}^\lhd}}
\newcommand*{\sourceport}[2]{\ensuremath{\mathit{#1}_{#2}^\rhd}}
\DeclareMathOperator{\domain}{domain}

%% file: head/acronyms.tex
\usepackage{acronym}
\acrodef{CGRA}{coarse-grained reconfigurable array}
\acrodef{TCPA}{Tightly Coupled Processor Array}
\acrodef{LDA}{linear dependence algorithm}
\acrodef{UDA}{uniform dependence algorithm}
\acrodef{PE}{processing element}
\acrodef{ILP}{integer linear program}
\acrodef{OIP}{orthogonal instruction processing}
\acrodef{CFG}{control flow graph}
\acrodef{RDG}{reduced dependence graph}
\acrodefindefinite{RDG}{an}{a}

%% file: sections/abstract.tex
Loop compilation for Tightly Coupled Processor Arrays (TCPAs), a class of massively parallel loop accelerators, entails solving NP-hard problems, yet depends on the loop bounds and number of available processing elements (PEs), parameters known only at runtime because of dynamic resource management and input sizes.
Therefore, this article proposes a two-phase approach called symbolic loop compilation:
At compile time, the necessary NP-complete problems are solved and the solutions compiled into a space-efficient \keyword{symbolic configuration}.
At runtime, a \keyword{concrete configuration} is generated from the symbolic configuration according to the parameters values.
We show that the latter phase, called \keyword{instantiation}, runs in polynomial time with its most complex step, program instantiation, not depending on the number of PEs.

As validation, we performed symbolic loop compilation on real-world loops and measured time and space requirements.
Our experiments confirm that a symbolic configuration is space-efficient and suited for systems with little memory---often, a symbolic configuration is smaller than a single concrete configuration---and that program instantiation scales well with the number of PEs---for example, when instantiating a symbolic configuration of a matrix-matrix multiplication, the execution time is similar for $4\times 4$ and $32\times 32$ PEs.

%% file: sections/introduction.tex
\section{Introduction}

\acp{TCPA}~\cite{hannig2014} are loop accelerators with the goal to be energy efficient by offering \emph{comprehensive} loop acceleration, meaning
they handle all parts of loop execution: computation, control, and communication.
For this purpose, \acp{TCPA} have a grid of numerous, simple \acp{PE} to exploit task- (multiple loops in parallel), loop- (multiple parts of a loop in parallel), iteration- (multiple subsequent iterations in parallel), and instruction-level parallelism;
they have global controllers to centrally compute control flow and unburden the \acp{PE},
a circuit-switched interconnect network to locally communicate intermediate data,
and I/O buffers with address generators to autonomously stream I/O data only at the borders.

Synchronization and efficient utilization of these components rely on a parallelizing compiler, in particular cycle-accurate scheduling of operations as well as high-quality register allocation and routing, all of which are NP-complete problems.
Because of this tight synchronization, the components require distinct programs and configuration data for any distinct combination of loop bounds and number of allocated \acp{PE}, but these two parameters are in general unknown a priori.
The number of allocated \acp{PE} is unknown because multiple applications may \emph{dynamically} allocate regions of \acp{PE} sized in accordance with, for example, non-functional properties such as latency and energy consumption.
We face a conundrum:
Both programs and configuration data must be generated at runtime despite the NP-complete problems compilation involves.
This renders just-in-time compilation unsuitable.
Instead, we propose to split compilation into two phases, as illustrated in Figure~\ref{fig:compile-flow}:
\begin{enumerate}
  \item \keyword{Symbolic mapping} (Section~\ref{sec:symbolic-compilation}) is performed off-line and solves the involved NP-complete problems, generating a \keyword{symbolic configuration}.
    A symbolic configuration is a novel compact representation of configurations parameterized in the loop bounds and number of \acp{PE}.
    Here, we contribute the first solution to allocating and representing routes on the interconnect network (Section~\ref{sec:routing}) despite not yet knowing the number of allocated \acp{PE}.
    The \ac{PE} programs are represented symbolically and compactly by a polyhedral syntax tree \cite{witterauf2019}.

  \item \keyword{Instantiation} (Section~\ref{sec:instantiation}) is performed once the parameter values are known and generates a concrete configuration from a symbolic configuration.
  In particular, we show for the first time how to instantiate \ac{PE} programs from a polyhedral syntax tree and that it is possible to instantiate them in polynomial time independently of the number of \acp{PE} (Sections \ref{sec:control-flow-analysis}--\ref{sec:program-generation}).
\end{enumerate}

In Section~\ref{sec:experiments}, we present experimental results showing the time and space efficiency of this hybrid compilation approach for a number of real-world loop programs---but first, we distinguish our work from related approaches to loop parallelization.

%% file: sections/related-work.tex
\section{Related Work}

Loop acceleration is a wide field of research; as for related work, we are interested in two aspects:
How are loops mapped to architectures similar to \acp{TCPA}, and what other \emph{symbolic} compilation approaches for parallelizing loops have been proposed?

\begin{figure}
    \centering
    \input{figures/tikz/compile-flow-slanted}
    \caption{Hybrid compilation flow for symbolically compiling loop programs for \acp{TCPA}.
      It is divided into symbolic mapping, performed offline, and instantiation, in general performed at runtime.}
    \label{fig:compile-flow}
  \end{figure}
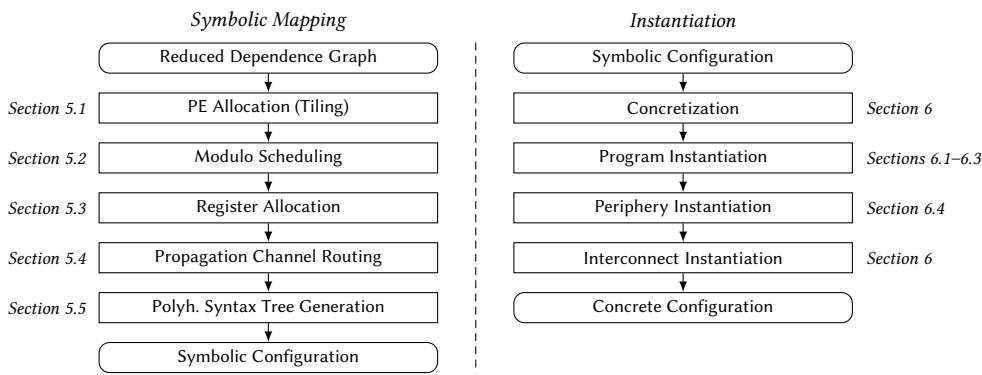
  
\subsection{Loop mapping on similar architectures}

Like \acp{TCPA}, the following classes of architectures are characterized by a grid of processing elements:
massively parallel processor arrays (MPPAs), coarse-grain reconfigurable arrays (CGRAs), and systolic neural network accelerators.

MPPAs are aimed at accelerating entire applications, not only loops;
consequently, the \acp{PE} of an MPPA are more complex, close to a general-purpose CPU, and usually communicate via shared memory and message passing.
Because cycle-accurate synchronization is not necessary then, an MPPA allows for more traditional compilation flows, such as the Kalray~MPPA-256~\cite{kalray2013}, which is programmed using OpenCL or POSIX multi-threading with automatic loop parallelization via OpenMP.
Another example is the KiloCore~\cite{kilocore2017}, which is programmed using C++ or assembly, but requires 
manual parallelization.
While the compilation flow proposed in this article only applies to a certain class of loops (see Section~\ref{sec:mapping-loops-to-tcpas}), they are automatically and comprehensively parallelized up to the loop level.
This includes the generation of programs and configuration data of tightly synchronized components dedicated to energy-efficient loop acceleration, whereas on MPPAs, the corresponding functionalities are usually part of the \ac{PE} programs.

CGRAs, on the other hand, only accelerate the kernel of a loop, that is, the body of the innermost level of a loop nest.
Their processing elements are usually simple, reconfigurable functional units interconnected by a circuit-switched network.
We refer to the survey \cite{wijtvliet2016} for details on CGRA architectures.
Simply put, loops are mapped onto a CGRA by embedding the data flow graph of the loop body onto the CGRA (nodes onto processing elements, edges onto interconnections).
While this type of mapping allows for arbitrary loop bodies, it not only disregards the vast parallelism offered by the regularity of a loop, but leaves loop control and I/O communication to the host CPU, making acceleration less autonomous and thus potentially less useful.
The generation of corresponding parallelized loop control and I/O code is usually not addressed in the respective papers.
By contrast, we offer a comprehensive approach to loop acceleration.

Finally, systolic neural network accelerators have very special-purpose compilation flows, starting from a neural network description, not a loop program.
They hence have a narrower scope than the compilation flow proposed in this article.
(Note that our compilation flow does support neural networks if formulated as a loop program.)

\subsection{Other symbolic loop compilation approaches}

%An ultimate benefit and advantage of TCPA architectures and the presented symbolic compilation technique is born on the insight that tiles of loop iterations must be processed locally inside each PE as much as possible. Different to many CGRA approaches that unnecessarily unfold the regular computations encountered in loop nests to create irregular data flow graphs that are subsequently mapped to irregular PE programs, the PE register structures of a TCPA have been   optimized particularly to avoid any non-necessary transfer of data.  In order to store intermediate data in the hardware of a PE, its particular register structure needs to be explained.  

Several works investigate the generation of loop code for a number of processors unknown at compile time.
DynTile~\cite{hartono2010} and D-Tiling~\cite{kim2009} target general-purpose multi-cores;
Kong et al.~\cite{kong2013} generate vectorized code for cores supporting SIMD processing;
Konstantinidis et al.~\cite{konstantinidis2014} generate parallelized code for GPUs.
However, none of these approaches apply to \acp{TCPA} because the target architectures do neither rely on cycle-accurate synchronization of components nor require \ac{PE}-specific compact programs (see Section~\ref{sec:control-flow-analysis}) to save space and keep instruction memories small.

These approaches also have in common that they finish code generation at compile time.
In contrast, the speculative loop optimizer Appollo offers two approaches for runtime code generation: code skeletons~\cite{jimborean2014} and, more recently, code bones~\cite{caamano2016}.
While these get assembled at runtime, similar to the instantiation phase described in this article, they lack the capability to represent instructions parameterized in the current iteration, which is necessary for generating modulo-scheduled programs in the compact manner required by \acp{TCPA} (see Section~\ref{sec:polyhedral-syntax-tree}).

Finally, while a symbolic configuration sounds similar to a template (static content mixed with placeholders that are replaced later), a symbolic configuration represents differently \emph{structured} configurations.
Next, before describing our hybrid compilation flow in detail, we discuss the fundamentals of \acp{TCPA} and how we model loops using reduced dependence graphs.

%% file: figures/tikz/compile-flow-slanted.tex
\begin{tikzpicture}[
  font=\footnotesize\sffamily,
  annotation/.style={align=left, font=\footnotesize\itshape},
  compile step/.style={rectangle, draw, minimum width=4.5cm, minimum height=0.4cm, inner sep=0pt, align=center},
  compile data/.style={rectangle, rounded corners=4pt, draw, minimum width=4.5cm, minimum height=0.4cm, inner sep=0pt},
  compile flow/.style={->, >=latex},
  node distance=0.25cm and 1cm,
  header/.style={font=\itshape\small, inner sep=0pt},
]
  % symbolic compilation

  \node[compile data, xshift=0.75cm] (loop) at (-0.25, 0) {Reduced Dependence Graph};
  \node[header, anchor=base, above=0.15cm of loop] {Symbolic Mapping};

  \node[compile step,below=of loop] (tiling) {PE Allocation (Tiling)}; 
  \node[compile step,below=of tiling] (scheduling) {Modulo Scheduling};
  \node[compile step,below=of scheduling] (register allocation) {Register Allocation};
  \node[compile step,below=of register allocation] (channel routing) {Propagation Channel Routing};
  \node[compile step,below=of channel routing] (polyhedral syntax tree) {Polyh.\ Syntax Tree Generation};
  \node[compile data,below=of polyhedral syntax tree] (symbolic configuration) {Symbolic Configuration};

  % instantiation

  \node[compile data,right=of loop] (symbolic configuration 2) {Symbolic Configuration}; 

  \node[header, anchor=base, above=0.2cm of symbolic configuration 2] {Instantiation};

  \node[compile step,below=of symbolic configuration 2] (control flow) {Concretization}; 
  \node[compile step,below=of control flow] (program generation) {Program Instantiation}; 
  \node[compile step,below=of program generation] (control signal allocation) {Periphery Instantiation}; 
  \node[compile step,below=of control signal allocation] (io configuration) {Interconnect Instantiation}; 
  \node[compile data,below=of io configuration] (binary) {Concrete Configuration}; 

  \path[compile flow]
    (loop) edge (tiling)
    (tiling) edge (scheduling)
    (scheduling) edge (register allocation)
    (register allocation) edge (channel routing)
    (channel routing) edge (polyhedral syntax tree)
    (polyhedral syntax tree) edge (symbolic configuration);
  \path[compile flow]
    (symbolic configuration 2) edge (control flow)
    (control flow) edge (program generation)
    (program generation) edge (control signal allocation)
    (control signal allocation) edge (io configuration)
    (io configuration) edge (binary);

  \draw[densely dashed]
    ($(loop.north east)!.5!(symbolic configuration 2.north west)$) coordinate (here)
    -- (symbolic configuration.south -| here);

  \begin{scope}[
    every node/.style={font=\footnotesize\itshape}
  ]
    \node[left=2pt of tiling] {Section~\ref{sec:symbolic-tiling}};
    \node[left=2pt of scheduling] {Section~\ref{sec:scheduling}};
    \node[left=2pt of register allocation] {Section~\ref{sec:register-allocation}};
    \node[left=2pt of channel routing] {Section~\ref{sec:routing}};
    \node[left=2pt of polyhedral syntax tree] {Section~\ref{sec:polyhedral-syntax-tree}};

    \node[right=2pt of control flow] {Section~\ref{sec:instantiation}};
    \node[right=2pt of program generation] {Sections~\ref{sec:control-flow-analysis}--\ref{sec:program-generation}};
    \node[right=2pt of control signal allocation] {Section~\ref{sec:io-access-generation}};
    \node[right=2pt of io configuration] {Section~\ref{sec:instantiation}};
  \end{scope}
\end{tikzpicture}

%% file: sections/tcpas.tex
\section{Tightly coupled processor arrays}
\label{sec:tcpas}

\begin{figure}
  \centering
  \input{figures/tikz/tcpa-block-diagram}
  \caption{%
    A Tightly Coupled Processor Array contains a grid of processing elements (PEs) interconnected using a circuit-switched network with two independent layers, data (solid lines) and control (dotted lines).
    Each PE is surrounded by an interconnect wrapper, whose internal architecture is shown on the right side for the data layer.
    Additionally, there are two kinds of peripheral components:
    I/O buffers, each providing one border with individual memory banks, and global controllers, one per corner, each orchestrating the parallel execution of one loop program.
    The figure shows two simultaneously running loop programs, indicated by the two colors.
  }
  \label{fig:tcpa-block-diagram}
\end{figure}
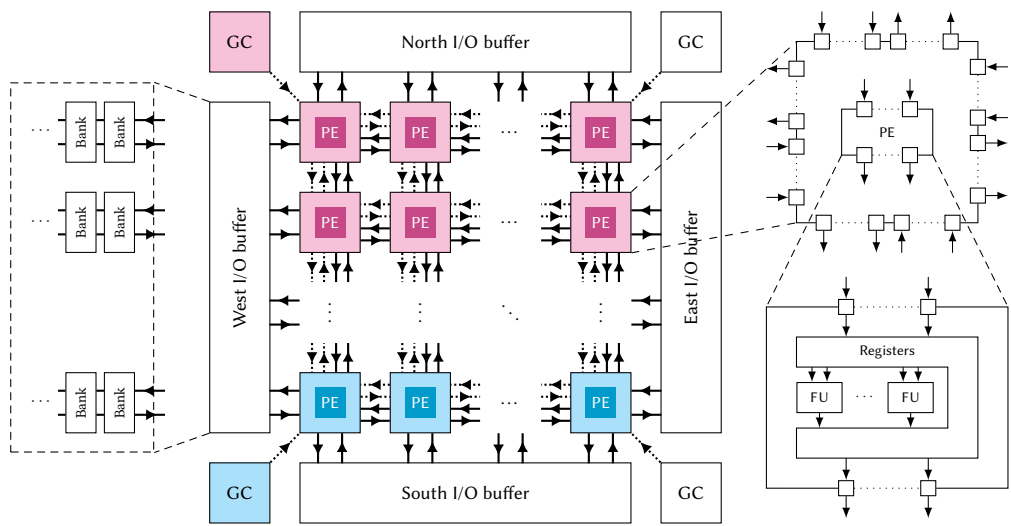

\acp{TCPA} are massively parallel loop accelerators highly configurable at synthesis time, featuring a grid of simple programmable \keyword{processing elements} (PEs)\acused{PE} \cite{kissler2006,hannig2014} interconnected by a circuit-switched network.
As illustrated in Figure~\ref{fig:tcpa-block-diagram}, the processor array is surrounded by \keyword{I/O buffers} on all four borders, responsible for decoupling data streaming from the rest of the system.
Finally, in each of the corners, there is a so-called \keyword{global controller}, responsible for synchronizing the parallel execution of a loop nest on a rectangular region of \acp{PE}.

\subsection{Interconnect network}

Instead of accessing shared memory, loop-carried dependences are communicated locally between processors to save energy.
Each \ac{PE} is embedded into a so-called interconnect wrapper that acts as a switch between its four neighbors and the \ac{PE}.
For that, each wrapper provides two independent layers, data and control, each with an individually configurable number of both input and output ports in the four cardinal directions, called wrapper ports, while the contained \ac{PE} provides input and output ports called \ac{PE} ports.
A $\mathit{port}$ is a triple $(\mathit{location}, \mathit{orientation}, i)$ where $i\in\mathbb{N}_0$ is an index, $\mathit{location}$ is one of $\mathit{north}$, $\mathit{east}$, $\mathit{south}$, $\mathit{west}$, or $\mathit{pe}$, and $\mathit{orientation}$ specifies whether it is a $\mathit{sink}$ (wrapper output and PE input ports) or a $\mathit{source}$ (wrapper input and PE output ports).
As a shorthand notation, we use $\sourceport{location}{i}$ for source ports and $\sinkport{location}{i}$ for sink ports.

Ports of opposing orientation may be connected.
Between wrappers, each wrapper port is connected to its \keyword{sibling}, which is the port of both opposing cardinal direction and opposing orientation;
for example $\sinkport{south}{0}$ of a wrapper is connected to $\sourceport{north}{0}$ of the wrapper to its south.
At the borders of the \ac{TCPA}, the wrapper ports facing the border are connected to corresponding input and output ports of the I/O buffer.
Within a wrapper, two ports can be connected at runtime if they are made \keyword{adjacent} at synthesis time; two ports may be adjacent if they are of opposing orientation.
%Combining synthesis and runtime configurability allows a wide variety of topologies.

%Generating the runtime configuration of the data layer of the interconnect network for a given loop program is the topic of Sections~\ref{sec:routing} and \ref{sec:io-access-generation}.

\subsection{Processing elements}
\label{sec:processing-elements}

Each \ac{PE} houses two types of registers: data and control.
In this paper, we focus on the data registers, which have been specially designed to exploit the regularity of loops based on the insight that each \ac{PE} processes a neighborhood of loop iterations locally:
\begin{itemize}
  \item \keyword{General-purpose registers} $R_\mathit{rd}=\{ \texttt{rd0}, \texttt{rd1}, \ldots \}$ are conventional registers and used for storing intermediate values during a single loop iteration.
  \item \keyword{Feedback registers} $R_\mathit{fd}=\{ \texttt{fd0}, \texttt{fd1}, \ldots \}$ are rotating shift registers with reconfigurable depth used for storing intermediate values across multiple loop iterations (loop-carried dependences).
    Each feedback register manages separate read and write pointers that advance (and potentially wrap around) after each read or write independently.
  \item \keyword{Input registers} $R_\mathit{id}=\{ \texttt{id0}, \texttt{id1}, \ldots \}$ each allow read access to a FIFO connected to the corresponding \ac{PE} input port: \texttt{id0} to $\sinkport{pe}{0}$, \texttt{id1} to $\sinkport{pe}{1}$, and so on.
  \item \keyword{Output registers} $R_\mathit{od}=\{ \texttt{od0}, \texttt{od1}, \ldots \}$ each allow write access to the corresponding \ac{PE} output port: \texttt{od0} to \sourceport{pe}{0}, \texttt{od1} to \sourceport{pe}{1}, and so on.
\end{itemize}
%The number of registers $|R_\mathit{*d}|$ is set at synthesis time.
The registers are connected to a set of functional units (FUs) $\{ \mathit{fu}_1, \mathit{fu}_2, \ldots \}$ 
composed at synthesis time from a variety of types such as integer and floating-point ALUs, MAC units, and so on.
Each type offers its own instruction set, but most follow the three-register format;
for example, \texttt{add\,rd0\,fd3\,id1} means $\texttt{rd0}\gets\texttt{fd3}+\texttt{id1}$.
For convenience, each functional unit is assigned a unique identifier such as \texttt{alu0} or \texttt{mac1}.

To enable instruction-level parallelism, each FU executes an individual program, an architecture known as orthogonal instruction processing (OIP) that aims to reduce the overall program size compared to VLIW architectures \cite{brand2017}.
In OIP, each FU contains an individual branch unit, which computes the program counter of the next instruction.
Conditional branches may depend on flags, which are shared between FUs, and on control registers, allowing the synchronization of programs across FUs.
The number of targets per conditional branch can be chosen at synthesis time, but is usually two.
Section~\ref{sec:program-generation} introduces branch instructions in more detail.

%\begin{figure}
  %\centering
  %\input{figures/tikz/tcpa-pe-block-diagram}
  %\caption{Processing element.}
  %\label{fig:tcpa-block-diagram}
%\end{figure}

%Generating \ac{PE} programs is the most complex part of \ac{TCPA} compilation and spans Sections~\ref{sec:scheduling}, \ref{sec:register-allocation}, \ref{sec:polyhedral-syntax-tree}, \ref{sec:control-flow-analysis}, \ref{sec:control-signal-allocation}, and \ref{sec:program-generation}.

\subsection{Periphery}

%Important for configuration data generation is also the periphery of a TCPA.
There are two types of supporting components:
four I/O buffers, one on each border, and up to four global controllers, one in each corner.
Each I/O buffer is flexibly connected to the respective border \acp{PE}, but the details irrelevant for this article, so we assume each border \ac{PE} is connected to an individual set of memory banks via its wrapper ports (compare Figure~\ref{fig:tcpa-block-diagram}).
Each memory bank contains a reconfigurable \keyword{address generator} for providing affine address sequences, unburdening the connected \ac{PE}.
%Their configuration is covered in Section~\ref{sec:io-access-generation}.
Each global controller orchestrates a single running loop application by generating control signals according to its iteration space and control flow, which are then propagated from \ac{PE} to \ac{PE} using the control layer of the interconnect network.
%Allocating control signals and configuring a global controller is discussed in Section~\ref{sec:control-signal-allocation}.

%\begin{figure}
  %\centering
  %\input{figures/tikz/tcpa-icn-wrapper}
  %\caption{%
    %Physical layout of an interconnect wrapper.
  %}
  %\label{fig:tcpa-icn-wrapper}
%\end{figure}

%\begin{figure}
  %\centering
  %\input{figures/tikz/icn-wrapper-as-switch}
  %\caption{%
    %An interconnect wrapper acts as a switch between the PE and the four neighbors.
  %}
  %\label{fig:tcpa-icn-switch}
%\end{figure}

%% file: figures/tikz/tcpa-block-diagram.tex
\tikzset{%
  tight background,
  tcpa pe/.style={
    draw=black,
    minimum size=1cm,
    inner sep=0pt,
  },
  tcpa pe inlay/.style={
    fill=gray,
    text=white,
    font=\sffamily\small,
    minimum size=0.5cm,
    inner sep=0pt,
  },
  tcpa periphery/.style={
    draw=black,
    text=black,
    font=\sffamily,
    inner sep=0pt,
  },
  tcpa gc/.style={
    tcpa periphery,
    minimum size=1cm,
  },
  tcpa memory/.style={
    tcpa periphery,
  },
  tcpa memory ns/.style={
    tcpa memory,
    minimum height=1cm,
    minimum width=5.5cm,
  },
  tcpa memory we/.style={
    tcpa memory,
    minimum height=1cm,
    minimum width=5.5cm,
    rotate=90,
  },
  tcpa wire/.style={
    thick,
    ->, >=latex,
  },
  tcpa memory index/.style={
    circle,
    font=\sffamily\scriptsize,
    inner sep=1pt,
    draw=black,
    text=black,
  },
  tcpa memory dots/.style={
    font=\sffamily\scriptsize,
    inner sep=1pt,
  },
}

\colorlet{tcpappA}{cyan}
\colorlet{tcpappB}{magenta}
\newcommand*{\inlaycolor}[1]{#1!80!black}
\newcommand*{\blockcolor}[1]{#1!30}

\begin{tikzpicture}[
  scale=0.8,
  transform shape,
  tcpa signal/.style={
    thick
  },
  tcpa data/.style={
    thick,
    decoration={
      markings,
      mark=at position 0.75 with {\arrow{Latex[length=1.5mm, width=1.5mm]}}
    },
    postaction={decorate}
  },
  tcpa control/.style={
    thick,
    densely dotted,
    decoration={
      markings,
      mark=at position 0.70 with {\arrow{Latex[length=1.5mm, width=1.5mm]}}
    },
    postaction={decorate}
  },
]
  % processor array
  \begin{scope}[shift={(-2.25, 2.25)}]
    % hidden PE nodes
    \foreach \y in {0,1,2,3} {
      \foreach \x in {0,1,2,3} {
        \node[tcpa pe, fill=none, draw=none] (pe-\y-\x) at ($1.5*(\x, 0)-1.5*(0, \y)$) {};
      }
    }

    % actual PE nodes
    \foreach \y in {0,1,3} {
      \foreach \x in {0,1,3} {
        \ifthenelse{\y=3}{
          \node[tcpa pe, fill=\blockcolor{tcpappA}] at (pe-\y-\x) {};
          \node[tcpa pe inlay, fill=\inlaycolor{tcpappA}] at (pe-\y-\x) {PE};  
        }{
          \node[tcpa pe, fill=\blockcolor{tcpappB}] at (pe-\y-\x) {};
          \node[tcpa pe inlay, fill=\inlaycolor{tcpappB}] at (pe-\y-\x) {PE};  
        }
      }
    }

    % ellipses
    \foreach \y in {0,1,3} {
      \node at (pe-\y-2) {\dots};
    }
    \foreach \x in {0,1,3} {
      \node at ([yshift=3pt] pe-2-\x) {\vdots};
    }
    \node at ([yshift=3pt] pe-2-2) {$\ddots$};

    % horizontal signals
    \foreach [evaluate=\x as \nx using {int(\x+1)}] \x in {0,...,2} {
      \foreach \y in {0,1,3} {
        \path ([yshift=0.3cm]pe-\y-\nx.west) edge[tcpa control] ([yshift=0.3cm]pe-\y-\x.east);
        \path ([yshift=0.1cm]pe-\y-\x.east) edge[tcpa control] ([yshift=0.1cm]pe-\y-\nx.west);
        \path ([yshift=-0.1cm]pe-\y-\nx.west) edge[tcpa data] ([yshift=-0.1cm]pe-\y-\x.east);
        \path ([yshift=-0.3cm]pe-\y-\x.east) edge[tcpa data] ([yshift=-0.3cm]pe-\y-\nx.west);
      }
    }
    % vertical signals
    \foreach [evaluate=\y as \ny using {int(\y+1)}] \y in {0,...,2} {
      \foreach \x in {0,1,3} {
        \path ([xshift=0.3cm]pe-\ny-\x.north) edge[tcpa data] ([xshift=0.3cm]pe-\y-\x.south);
        \path ([xshift=0.1cm]pe-\y-\x.south) edge[tcpa data] ([xshift=0.1cm]pe-\ny-\x.north);
        \path ([xshift=-0.1cm]pe-\ny-\x.north) edge[tcpa control] ([xshift=-0.1cm]pe-\y-\x.south);
        \path ([xshift=-0.3cm]pe-\y-\x.south) edge[tcpa control] ([xshift=-0.3cm]pe-\ny-\x.north);
      }
    }
  \end{scope}

  % global controllers
  \begin{scope}
    \begin{scope}
      \node[tcpa gc, fill=\blockcolor{tcpappB}] (gc-nw) at (-3.75, 3.75) {GC};
      \node[tcpa gc] (gc-ne) at (3.75, 3.75) {GC};
      \node[tcpa gc, fill=\blockcolor{tcpappA}] (gc-sw) at (-3.75, -3.75) {GC};
      \node[tcpa gc] (gc-se) at (3.75, -3.75) {GC};
    \end{scope}
    \path (gc-nw) edge[tcpa control] (pe-0-0);
    \path (gc-ne) edge[tcpa control] (pe-0-3);
    \path (gc-se) edge[tcpa control] (pe-3-3);
    \path (gc-sw) edge[tcpa control] (pe-3-0);
  \end{scope}

  % memories
  \begin{scope}
    \node[tcpa memory ns] (memory-north) at (0, 3.75) {North I/O buffer};
    \node[tcpa memory ns] (memory-south) at (0, -3.75) {South I/O buffer};
    \node[tcpa memory we] (memory-west) at (-3.75, 0) {West I/O buffer};
    \node[tcpa memory we] (memory-east) at (3.75, 0) {East I/O buffer};

    % Memories -> PEs
    \foreach \x in {0,1,2,3} {
      \draw[tcpa data] ([xshift=0.2cm] pe-0-\x.north) -- ++(up:0.5);
      \draw[tcpa data] ([xshift=-0.2cm] pe-0-\x.north) ++(up:0.5) -- ([xshift=-0.2cm] pe-0-\x.north);
      \draw[tcpa data] ([xshift=-0.2cm] pe-3-\x.south) -- ++(down:0.5);
      \draw[tcpa data] ([xshift=0.2cm] pe-3-\x.south) ++(down:0.5) -- ([xshift=0.2cm] pe-3-\x.south);
    }
    \foreach \y in {0,1,2,3} {
      \draw[tcpa data] ([yshift=0.2cm] pe-\y-0.west) -- ++(left:0.5);
      \draw[tcpa data] ([yshift=-0.2cm] pe-\y-0.west) ++(left:0.5) -- ([yshift=-0.2cm] pe-\y-0.west);
      \draw[tcpa data] ([yshift=-0.2cm] pe-\y-3.east) -- ++(right:0.5);
      \draw[tcpa data] ([yshift=0.2cm] pe-\y-3.east) ++(right:0.5) -- ([yshift=0.2cm] pe-\y-3.east);
    }

    \foreach \i in {0, 1, 3} {
      \node[minimum width=0.5cm, minimum height=1cm, draw] 
        (bank-west-\i-0) at ([xshift=-2cm] pe-\i-0.center -| memory-west.west) {};
      \node[rotate=90, font=\footnotesize] at (bank-west-\i-0) {Bank};  
      \node[minimum width=0.5cm, minimum height=1cm, draw, left=0.125cm of bank-west-\i-0] (bank-west-\i-1) {};
      \node[rotate=90, font=\footnotesize] at (bank-west-\i-1) {Bank};
      \node[left=0.2cm of bank-west-\i-1, inner sep=0pt, font=\footnotesize] (bank-west-\i-dots) {$\cdots$};

      \draw[tcpa data] ([yshift=-0.2cm]bank-west-\i-0.east) -- ++(right:0.5cm);
      \draw[tcpa data] ([yshift=0.2cm]bank-west-\i-0.east) ++(right:0.5cm) -- ++(left:0.5cm);
      \draw[thick] ([yshift=-0.2cm]bank-west-\i-1.east) -- ([yshift=-0.2cm]bank-west-\i-0.west);
      \draw[thick] ([yshift=0.2cm]bank-west-\i-1.east) -- ([yshift=0.2cm]bank-west-\i-0.west);
      \draw[thick] ([yshift=-0.2cm]bank-west-\i-1.west) -- ++(left:0.125cm);
      \draw[thick] ([yshift=0.2cm]bank-west-\i-1.west) -- ++(left:0.125cm);
  }

    \node[draw, inner sep=0.25cm, densely dashed, fit={(bank-west-0-0.north east) (bank-west-3-0.south east) (bank-west-3-dots.south west)}] (memory-west-exploded) {};

    \draw[densely dashed]
      (memory-west.north east) -- (memory-west-exploded.north east)
      (memory-west.north west) -- (memory-west-exploded.south east);
  \end{scope}

  \begin{scope}
    \node[rectangle, thin, dashed, minimum size=3cm, right=1.75cm of memory-east, anchor=north west, yshift=1cm] (wrapper-exploded) {};
    \node[rectangle, minimum width=1.5cm, minimum height=0.75cm, font=\footnotesize\sffamily] (pe-exploded) at (wrapper-exploded) {PE};

    \newcommand\offset{}
    \foreach \direction in {north,east,south,west} {
      \newcommand\orientationa{input}
      \newcommand\orientationb{output}

        \ifthenelse{\equal{\direction}{north}}{
        \renewcommand\offset{up:0.5cm}
        \coordinate (corner a) at (wrapper-exploded.north west);
        \coordinate (corner b) at (wrapper-exploded.north east);
      }{}
      \ifthenelse{\equal{\direction}{south}}{
        \renewcommand\offset{down:0.5cm}
        \coordinate (corner b) at (wrapper-exploded.south west);
        \coordinate (corner a) at (wrapper-exploded.south east);
        \renewcommand\orientationa{output}
        \renewcommand\orientationb{input}
      }{}
      \ifthenelse{\equal{\direction}{west}}{
        \renewcommand\offset{left:0.5cm}
        \coordinate (corner b) at (wrapper-exploded.north west);
        \coordinate (corner a) at (wrapper-exploded.south west);
        \renewcommand\orientationa{output}
        \renewcommand\orientationb{input}
      }{}
      \ifthenelse{\equal{\direction}{east}}{
        \renewcommand\offset{right:0.5cm}
        \coordinate (corner a) at (wrapper-exploded.north east);
        \coordinate (corner b) at (wrapper-exploded.south east);
      }{}
  
      \node[rectangle, draw, minimum size=0.25cm, inner sep=0pt]
        (port-\direction-0) at ($(corner a)!.14!(corner b)$) {};
      \node[rectangle, draw, minimum size=0.25cm, inner sep=0pt]
        (port-\direction-1) at ($(corner a)!.42!(corner b)$) {};
      \node[rectangle, draw, minimum size=0.25cm, inner sep=0pt]
        (port-\direction-2) at ($(corner a)!.56!(corner b)$) {};
      \node[rectangle, draw, minimum size=0.25cm, inner sep=0pt]
        (port-\direction-3) at ($(corner a)!.85!(corner b)$) {};

      \draw[->,>=latex]
        (port-\direction-2) -- ++(\offset);
      \draw[->,>=latex]
        (port-\direction-3) -- ++(\offset);
      \draw[<-,>=latex]
        (port-\direction-0) -- ++(\offset);
      \draw[<-,>=latex]
        (port-\direction-1) -- ++(\offset);

      \draw[dotted]
        (port-\direction-0) -- (port-\direction-1)
        (port-\direction-2) -- (port-\direction-3);
      \draw
        (port-\direction-1) -- (port-\direction-2);
    }

    \draw
      (port-west-3) |- (port-north-0)
      (port-north-3) -| (port-east-0)
      (port-east-3) |- (port-south-0)
      (port-south-3) -| (port-west-0)
      ;

    \node[rectangle, draw, minimum size=0.25cm, inner sep=0pt]
      (port-pe-input-1) at ($(pe-exploded.north west)!.25!(pe-exploded.north east)$) {};
    \node[rectangle, draw, minimum size=0.25cm, inner sep=0pt]
      (port-pe-input-n) at ($(pe-exploded.north west)!.75!(pe-exploded.north east)$) {};
    \node[rectangle, draw, minimum size=0.25cm, inner sep=0pt]
      (port-pe-output-1) at ($(pe-exploded.south west)!.25!(pe-exploded.south east)$) {};
    \node[rectangle, draw, minimum size=0.25cm, inner sep=0pt]
      (port-pe-output-n) at ($(pe-exploded.south west)!.75!(pe-exploded.south east)$) {};

    \draw[<-, >=latex]
      (port-pe-input-1) -- ++(up:0.5cm);
    \draw[<-, >=latex]
      (port-pe-input-n) -- ++(up:0.5cm);
    \draw[->, >=latex]
      (port-pe-output-1) -- ++(down:0.5cm);
    \draw[->, >=latex]
      (port-pe-output-n) -- ++(down:0.5cm);

    \draw[dotted]  
      (port-pe-input-1) -- (port-pe-input-n)
      (port-pe-output-1) -- (port-pe-output-n);
    \draw
      (port-pe-output-1) -- (pe-exploded.south west) |- (port-pe-input-1);
    \draw
      (port-pe-input-n) -- (pe-exploded.north east) |- (port-pe-output-n);

    \draw[dashed]
      (pe-1-3.north east) -- (wrapper-exploded.north west);
    \draw[dashed]
      (pe-1-3.south east) -- (wrapper-exploded.south west);

  \end{scope}

  \begin{scope}[font=\sffamily\footnotesize]
    \coordinate (pe-exploded-sw) at (pe-exploded.south west);
    \coordinate (pe-exploded-se) at (pe-exploded.south east);
    \node[thin, dashed, rectangle, minimum width=4cm, minimum height=3cm, below=1.4cm of wrapper-exploded] (pe-exploded) {};
    
    \draw[densely dashed] (pe-exploded-sw) -- (pe-exploded.north west);
    \draw[densely dashed] (pe-exploded-se) -- (pe-exploded.north east);

    \node[rectangle, draw, minimum size=0.25cm, inner sep=0pt]
      (port-pe-input-1) at ($(pe-exploded.north west)!.33!(pe-exploded.north east)$) {};
    \node[rectangle, draw, minimum size=0.25cm, inner sep=0pt]
      (port-pe-input-n) at ($(pe-exploded.north west)!.67!(pe-exploded.north east)$) {};
      \node[rectangle, draw, minimum size=0.25cm, inner sep=0pt]
      (port-pe-output-1) at ($(pe-exploded.south west)!.33!(pe-exploded.south east)$) {};
    \node[rectangle, draw, minimum size=0.25cm, inner sep=0pt]
      (port-pe-output-n) at ($(pe-exploded.south west)!.67!(pe-exploded.south east)$) {};

    \draw[<-, >=latex]
      (port-pe-input-1) -- ++(up:0.5cm);
    \draw[<-, >=latex]
      (port-pe-input-n) -- ++(up:0.5cm);
    \draw[->, >=latex]
      (port-pe-output-1) -- ++(down:0.5cm);
    \draw[->, >=latex]
      (port-pe-output-n) -- ++(down:0.5cm);

    \draw[dotted]  
      (port-pe-input-1) -- (port-pe-input-n)
      (port-pe-output-1) -- (port-pe-output-n);
    \draw
      (port-pe-output-1) -- (pe-exploded.south west) |- (port-pe-input-1);
    \draw
      (port-pe-input-n) -- (pe-exploded.north east) |- (port-pe-output-n);

    \draw
      ([shift={(0.5, -0.5)}] pe-exploded.north west) coordinate (outer-nw)
      -- ([shift={(-0.5, -0.5)}] pe-exploded.north east) coordinate (outer-ne)
      -- ([shift={(-0.5, 0.5)}] pe-exploded.south east) coordinate (outer-se)
      -- ([shift={(0.5, 0.5)}] pe-exploded.south west) coordinate (outer-sw)
      -- ++(up:0.5cm) coordinate (sw)
      -- ([shift={(-1, 1)}] pe-exploded.south east) coordinate (se)
      -- ([shift={(-1, -1)}] pe-exploded.north east) coordinate (ne)
      -- ([shift={(0.5, -1)}] pe-exploded.north west) coordinate (nw)
      -- cycle
      ;
    \node at ([yshift=-0.75cm] pe-exploded.north) {Registers};

    \draw[->, >=latex] (port-pe-input-1) -- (port-pe-input-1 |- outer-nw);
    \draw[->, >=latex] (port-pe-input-n) -- (port-pe-input-n |- outer-nw);
    \draw[<-, >=latex] (port-pe-output-1) -- (port-pe-output-1 |- outer-sw);
    \draw[<-, >=latex] (port-pe-output-n) -- (port-pe-output-n |- outer-sw);

    \node[minimum width=0.75cm, minimum height=0.5cm, draw, anchor=north west] (fu-0) at ([yshift=-0.25cm] nw) {FU};
    \node[minimum width=0.75cm, minimum height=0.5cm, draw, right=0.75cm of fu-0] (fu-n) {FU};
    \draw[->, >=latex] (fu-0.south) -- (fu-0.south |- sw);
    \draw[->, >=latex] (fu-n.south) -- (fu-n.south |- sw);
    \draw[<-, >=latex] ([xshift=-0.125cm] fu-0.north) coordinate (temp) -- (temp |- nw);
    \draw[<-, >=latex] ([xshift=0.125cm] fu-0.north) coordinate (temp) -- (temp |- nw);
    \draw[<-, >=latex] ([xshift=-0.125cm] fu-n.north) coordinate (temp) -- (temp |- nw);
    \draw[<-, >=latex] ([xshift=0.125cm] fu-n.north) coordinate (temp) -- (temp |- nw);

    \node[font=\scriptsize\sffamily] at ($(fu-0.east)!.5!(fu-n.west)$) {$\cdots$};
  \end{scope}

  %\begin{scope}[xshift=7.25cm, yshift=-4.4cm]
    %\draw[tcpa data] (0, 0.5) -- ++(right:0.75cm) node[right] {Data}; 
    %\draw[tcpa control] (0, 1) -- ++(right:0.75cm) node[right] {Control}; 
  %\end{scope}

\end{tikzpicture}

%% file: sections/prerequisites.tex
\section{Loops, the Polyhedral Model, and Reduced Dependence Graphs}
\label{sec:mapping-loops-to-tcpas}

As shown in Figure~\ref{fig:compile-flow}, we assume that compilation for a \ac{TCPA} starts from a loop program given as \iac{RDG}.
For transforming other loop representations, in particular sequential loops, into \acp{RDG}, we refer to the literature~\cite{feautrier1991}.

Reduced dependence graphs are closely tied to the polyhedral model, where the \keyword{iteration space} $\mathcal{I}$ of an $n$-dimensional loop nest is represented as a subset $\mathcal{I}$ of $\mathbb{Z}^n$, and each iteration is identified by a vector $\vec{I}\in\mathcal{I}$.
In particular, our loop model is based on \keyword{piecewise linear dependence algorithms} (PLAs) \cite{thiele1991}.
A PLA is a set of equations $S_i$ quantified over $\mathcal{I}$ that interrelate the \keyword{instances} of a set $X$ of affinely indexed variables, where the instance of $x\in X$ at index $\vec{I}$ is denoted $x[\vec{I}]$.
In this article, we assume the form
\begin{equation}\label{eq:pla}
  S_i\colon x_i[Q_i\vec{I}-\vec{d}_i] = \mathit{op}_i(x_{i, 1}[Q_{i,1}\vec{I}-\vec{d}_{i,1}], x_{i, 2}[Q_{i,2}\vec{I}-\vec{d}_{i,2}], \ldots) \textbf{ if } \vec{I}\in\mathcal{I}_i,
\end{equation}
where $x_i\in X$ is the variable with instances \keyword{defined} by $S_i$ as the result of operation $\mathit{op}_i$ and $x_{i, j} \in X$ are the variables with instances \keyword{used} by the definition.
Each variable is $m_i$-dimensionally indexed using an affine transform given by a matrix $Q\in\mathbb{Z}^{m_i\times n}$ and a vector $\vec{d}\in\mathbb{Z}^{m_i}$.
Which instances of $x_i$ are defined by an equation $S_i$ is restricted by its \keyword{condition space} $\mathcal{I}_i\subset\mathbb{Z}^n$.
Note that no instance of a variable $x\in X$ may be defined more than once by a PLA\footnote{This corresponds to the array single-assignment property.}.
We assume the iteration space $\mathcal{I}$ of a PLA to be the union of its condition spaces.

%We partition the set of variables $X$ into three subsets $X_\mathit{var}$, $X_\mathit{in}$, and $X_\mathit{out}$.
Without loss of generality, we use a restricted form of Equation~\eqref{eq:pla} that describes \keyword{uniform dependence algorithms with affinely indexed I/O variables}.
Here, each variable $x\in X$ satisfies exactly one of the following conditions:
(1) if instances of $x$ are only used, but none are defined (intuitively: $x$ only appears on the right-hand side), $x$ is an \keyword{input variable};
(2) if instances of $x$ are only defined, but none are used (intuitively: $x$ only appears on the left-hand side), $x$ is an \keyword{output variable}; and
(3) if all used instances of $x$ are defined, $x$ is an \keyword{internal variable}.
These conditions partition $X$ into the set of input variables $X_\mathit{in}$, the set of output variables $X_\mathit{out}$, and the set of internal variables $X_\mathit{var}$.\footnote{Any PLA may be transformed to allow for such a partitioning.}
Furthermore, internal variables may only be indexed uniformly\footnote{PLAs may be systematically transformed into this form by localization and embedding \cite{teich1991, thiele1989}.}:
When defining an internal variable $x_i\in X_\mathit{var}$, the matrix $Q_i$ must be the identity matrix and the vector $\vec{d}_i$ must be $\vec{0}$;
when using an internal variable $x_{i, j}\in X_\mathit{var}$, then $Q_{i, j}$ must be the identity matrix.
In the latter case, we call $\vec{d}_{i, j}$ the \keyword{dependence vector} of the \keyword{uniform dependence} between $x_{i, j}$ and $x_i$.

Note that input and output scalars are modeled as zero-dimensionally indexed input and output variables.
If the value $c$ of an input scalar $x\in X_\mathit{in}$ is known a priori, we call it a constant and refer to it by $c$, that is $x\coloneqq c$.

\begin{example}
Throughout this article, we use the following artificial, yet illustrative example that writes the first $N$ bits of an integer scalar $\mathit{in}$ into an array $\mathit{bits}$:
\begin{algorithmic}
\For {$0\leq i < N $} \Comment{iteration space is union of all condition spaces}
  \State $S_1\colon x[i] = \mathit{in}[]$ \textbf{if} $i=0$ \Comment{read scalar input $\mathit{in}$ in first iteration}
  \State $S_2\colon x[i] = y[i-1]$ \textbf{if} $i \geq 1$ \Comment{otherwise, use value shifted $i$ times}
  \State $S_3\colon y[i] = x[i] \texttt{\,shr\,} 1$ \Comment{shift right by constant 1 for next iteration}
  \State $S_4\colon \mathit{bits}[i] = x[i] \texttt{\,and\,} 1$ \Comment{extract bit and output into $\mathit{bits}$}
\EndFor
\end{algorithmic}
Note that $1\in X_\mathit{in}$ is a constant used in both $S_3$ and $S_4$.
\end{example}

PLAs prescribe neither place nor time of execution;
feasible execution orders are only implied by the dependences between equations.
A \keyword{reduced dependence graph} (RDG) is a directed graph $(V, E)$ that makes these explicit.
In the following, we rely on node annotations and edge annotations to structure all relevant information.
An annotation is a function $f$ that maps either a node $v\in V$ or an edge $e\in E$ to a value $z$ of arbitrary type.
To disambiguate function application from annotations, we write $z=f[v]$ to access an annotated value and $f[v]\gets z$ to annotate a value.

Given a uniform dependence algorithm with affinely indexed I/O variables, the corresponding \ac{RDG} is a directed multigraph $(V, E)$ with the following nodes:
\begin{itemize}
  \item An \keyword{operation node} $v$ for each equation $S_i$, annotated with the operation $\mathit{op}[v]\gets \mathit{op}_i$ and condition space $\mathcal{I}[v]\gets \mathcal{I}_i$.

  \item An \keyword{input node} $v$ for each $x_\mathit{in}\in X_\mathit{in}$ that is not a constant, annotated with the variable $x[v]\gets x_\mathit{in}$, and an \keyword{output node} for each $x_\mathit{out} \in X_\mathit{out}$, annotated with the variable $x[v]\gets x_\mathit{out}$.

  \item A \keyword{constant node} $v$ for each constant $c\in X_\mathit{in}$, annotated with its value $\annotate{c}{v} \gets c$.
\end{itemize}

\begin{figure}
  \centering
  \input{figures/tikz/rdg}
  \caption{Reduced dependence graph for the running example (bit extraction).
    Each rectangle is an operation node that represents an equation $S_i$ and is annotated with its sink variable $x_i$, its operation $\mathit{op}_i$ (in brackets) and its iteration-dependent condition space $\mathcal{I}_i$ (second line).
    Each rounded rectangle is an input or output node, and each circle is a constant node.
    Additionally, each operation node is annotated with the result of scheduling (Section~\ref{sec:scheduling}) in the form $\mathit{fu}$: [$\tau$] $\mathit{mnemo}$.
    Each edge represents a dependence and is annotated with the corresponding indexing function (here only shown for input and output edges and if $\vec{d}\neq\vec{0}$) and the its operand index (here either $A\coloneqq 1$ or $B\coloneqq 2$).
    The two dashed edges $e_4^1$ and $e_4^2$ are the result of splitting edge $e_4$ according to its dependence's processor displacements (Section~\ref{sec:symbolic-tiling}).
    Finally, each edge is annotated with its allocated register(s) (Section~\ref{sec:register-allocation}).
  }
  \label{fig:rdg}
\end{figure}
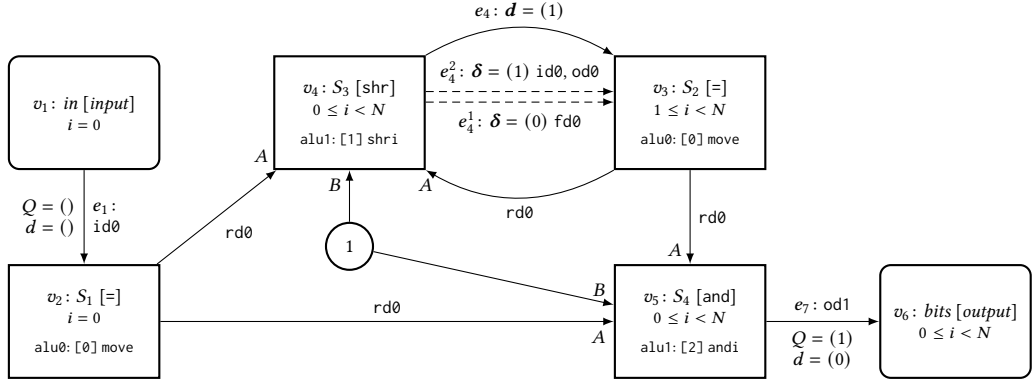

The set of edges $E$ contains the following edges:
\begin{itemize}
  \item If $w$ is an operation node representing equation $S_i$ and $v$ is an operation node representing $S_k$, a \keyword{dependence edge} $e=(v, w)$ from $v$ to $w$ is inserted for each $j$ where $x_k=x_{i, j}$ (that is, for each use of $x_k$ in $S_i$), annotated with dependence vector $\vec{d}[e]\gets \vec{d}_{i, j}$ and operand index $\annotate{\mathit{pos}}{e}\gets j$.

  \item If $w$ is an operation node representing equation $S_i$ and $v$ is the input node representing $x_\mathit{in}\in X_\mathit{in}$, an \keyword{input edge} $e=(v, w)$ is inserted for each $j$ where $x_\mathit{in}=x_{i, j}$ (that is, for each use of $x_\mathit{in}$ in $S_i$), annotated with the indexing function $\annotate{Q}{e} \gets Q_{i, j}$ and $\annotate{\vec{d}}{e} \gets \vec{d}_{i, j}$, as well as the input variable $\annotate{x}{e} \gets x_\mathit{in}$ and operand index $\annotate{\mathit{pos}}{e}\gets j$.

  \item If $w$ is an operation node representing equation $S_i$ and $v$ is a constant node representing value $c$, a \keyword{constant edge} $e=(v, w)$ is inserted for each $j$ where $c=x_{i, j}$ (that, is for each use of $c$ in $S_i$), annotated with the operand index $\annotate{\mathit{pos}}{e}\gets j$.

  \item If $v$ is an operation node representing $S_i$ and $w$ is the output node representing $x_i\in X_\mathit{out}$, an \keyword{output edge} is inserted, annotated with the indexing function $\annotate{Q}{e} \gets Q_i$ and $\annotate{\vec{d}}{e} \gets \vec{d}_i$, as well as the output variable $\annotate{x}{e} \gets x_i$.
\end{itemize}
Figure~\ref{fig:rdg} shows the \ac{RDG} of the running example, which we use as the basis in the next section to explain how to compile a loop program given as \iac{RDG} into a symbolic configuration.

%% file: figures/tikz/rdg.tex
  \begin{tikzpicture}[
    every node/.style={font=\footnotesize},
    rdg node/.style={draw, thick, inner sep=0pt, minimum height=0.5cm},
    rdg operation node/.style={rdg node, minimum width=2cm, minimum height=1.5cm, align=center},
    rdg input node/.style={rdg node, rounded corners=4pt, minimum width=2cm, minimum height=1.5cm, align=center},
    rdg output node/.style={rdg node, rounded corners=4pt, minimum width=2cm, minimum height=1.5cm, align=center},
    rdg constant node/.style={rdg node, circle, minimum size=0.625cm},
    rdg edge/.style={->, >=latex},
    rdg label/.style={font=\footnotesize},
    rdg annotation/.style={font=\scriptsize},
    node distance=1.25cm and 1.5cm,
    every loop/.style={min distance=1.5cm, in=70, out=110},
    tight background,
  ]
    \node[rdg input node] (input) {$v_1\colon \mathit{in}$ [\textit{input}]\\\scriptsize $i = 0$};
    \node[rdg operation node, below=of input] (s1) {$v_2\colon S_1$ [\texttt{=}]\\\scriptsize $i = 0$\\[3pt]\scriptsize \texttt{alu0}: \texttt{[0]} \texttt{move}};
    \node[rdg operation node, right=of input] (s3) {$v_4\colon S_3$ [\texttt{shr}]\\\scriptsize $0 \leq i < N$\\[3pt]\scriptsize \texttt{alu1}: \texttt{[1]} \texttt{shri}};
    \node[rdg operation node, right=2.5cm of s3] (s2) {$v_3\colon S_2$ [\texttt{=}]\\\scriptsize $1 \leq i < N$\\[3pt]\scriptsize \texttt{alu0}: \texttt{[0]} \texttt{move}};
    \node[rdg operation node, below=of s2] (s4) {$v_5\colon S_4$ [\texttt{and}]\\\scriptsize $0 \leq i < N$\\[3pt]\scriptsize \texttt{alu1}: \texttt{[2]} \texttt{andi}};
    \node[rdg constant node] (one) at ([yshift=1cm] s4 -| s3) {$1$};
    \node[rdg output node, right=of s4] (output) {$v_6\colon \mathit{bits}$ [\textit{output}]\\\scriptsize $0 \leq i < N$};

    \path[rdg edge]
      (one) edge node[rdg label, at end, below left] {$B$} (s3)
            edge node[rdg label, at end, above left] {$B$} (s4)
      (input) edge node[rdg label, right, align=left] {$e_1\colon$\\\texttt{id0}} node[rdg label, left, align=right] {$Q=()$\\$d=()$} (s1)
      (s1) edge node[rdg label, below right] {\texttt{rd0}} node[rdg label, at end, above left] {$A$} (s3)
           edge node[rdg label, above] {\texttt{rd0}} node[rdg label, at end, below left] {$A$} (s4)
      (s3.north east) edge[bend left] node[rdg label, above] {$e_4\colon \vec{d}=(1)$} (s2.north west)
      (s2.south west) edge[bend left] node[rdg label, below] {\texttt{rd0}} node[rdg label, at end, below] {$A$} (s3.south east)
      (s2) edge node[rdg label, right] {\texttt{rd0}} node[rdg label, at end, above left] {$A$} (s4)  
      (s4) edge node[rdg label, above] {$e_7\colon$\texttt{od1}} node[rdg label, below, align=center] {$Q=(1)$\\$d=(0)$} (output)
      ;

    \draw[rdg edge, densely dashed]
      ([yshift=4pt] s3.east) -- node[below] {$e_4^1\colon \vec{\delta}=(0)$\ \texttt{fd0}} ([yshift=4pt] s2.west);
    \draw[rdg edge, densely dashed]
      ([yshift=8pt] s3.east) -- node[above] {$e_4^2\colon \vec{\delta}=(1)$\ \texttt{id0}, \texttt{od0}} ([yshift=8pt] s2.west);

\end{tikzpicture}

%% file: sections/compile-time.tex
\section{Symbolic mapping}
\label{sec:symbolic-compilation}

Because the loop bounds and number of \acp{PE} are assumed to become known only at run time, mapping \iac{RDG} onto a \ac{TCPA} is split into two phases (see Figure~\ref{fig:compile-flow}):
Symbolic mapping, which front-loads as many steps as possible to be performed at compile time and produces a symbolic configuration, and instantiation, which generates a concrete configuration from it.
In this section, we discuss the former, consisting of the following steps:
\begin{enumerate}
  \item \textbf{Processor allocation} by tiling and \textbf{symbolic modulo scheduling} define a space-time mapping of the \ac{RDG}, assigning each operation node a time, a \ac{PE}, as well as a functional unit and corresponding instruction.
    Both tiling and modulo scheduling are well known, but we review them in Sections~\ref{sec:symbolic-tiling} and \ref{sec:scheduling}.

  \item \textbf{Register allocation} (Section~\ref{sec:register-allocation}) assigns registers to edges in the \ac{RDG} that store and communicate both intermediate results and I/O data.

  \item \textbf{Routing of propagation channels} (Section~\ref{sec:routing}) allocates routes on the interconnect network for propagating intermediate values according to the data dependences.

  \item \textbf{Generation of a polyhedral syntax tree}.
    This data structure is a compact representation of all programs that adhere to a given space-time mapping, register allocation, and channel routing.
    It is independent of runtime parameters such as loop bounds and number of available \acp{PE} in the sense that given any valid values of these parameters, the corresponding program can be generated from it.
    While polyhedral syntax trees were first introduced in \cite{witterauf2019}, we review them in Section~\ref{sec:polyhedral-syntax-tree}.
\end{enumerate}
These steps are performed at compile time because they contain NP-complete problems but heavily influence the quality (code size, number of registers, and so on) of the generated configuration and therefore benefit from not being time-constrained.
The final output of the symbolic mapping phase is a symbolic configuration, which retains parameters as symbols, but from which concrete configurations can be instantiated once the parameter values become known.

%% Allocation %%%%%%%%%%%%%%%%%%%%%%%%%%%%%%%%%%%%%%%%%%%%%%%%%%%%%%%%%%%%%%%%%%%%%%%%%%%%%%%%%%%%%

\subsection{Allocation of processing elements by tiling}
\label{sec:symbolic-tiling}

To distribute the loop iterations across \acp{PE} for execution, our compilation starts with \keyword{orthogonal tiling}:
partitioning a loop's iteration space into $t_1\times t_2\times \dots t_n$ rectangular tiles, each of size $p_1\times p_2\times \dots p_n$.
Assuming that at most two tile counts $t_r$ and $t_c$ are not 1, the tiles are mapped one-to-one to a rectangular region $t_r\times t_c$ of \acp{PE} on the \ac{TCPA}, with each \ac{PE} being assigned the execution of the iterations within one tile.

Mathematically, tiling decomposes an iteration space $\mathcal I$ into an intra-tile iteration space $\mathcal J$ and inter-tile iteration space $\mathcal K$~\cite{teich1993}:
\[
    \mathcal{I}\subseteq\mathcal{J}\oplus\mathcal{K}=\left\{(\vec{J}, \vec{K})^\T \mid \vec{J}\in\mathcal{J}, \vec{K}\in\mathcal{K} \right\},
\]
where we call $\mathcal{I}^*=\mathcal{J}\oplus\mathcal{K}\in\mathbb{Z}^{2n}$ the \keyword{tiled iteration space}.
We assume a rectangular inter-tile iteration space, which describes the set of tiles:
\[
    \mathcal{K}=\left\{ \vec{K}=(k_1, k_2, \ldots, k_n)^\T \mid 0\leq k_i< t_i \right\}
\]
Likewise, the intra-tile iteration space describes the set of iterations within every tile:
\[
    \mathcal{J}=\left\{ \vec{J}=(j_1, j_2, \ldots, j_n)^\T \mid 0\leq j_i< p_i \right\}.
\]
Explained informally, tiling transforms an $n$-dimensional loop into a $2n$-dimensional loop where $n$ dimensions iterate over the tiles and $n$ dimensions iterate over the iterations within a tile.
Since tiling doubles the dimension of a given loop nest, condition spaces $\annotate{\mathcal{I}}{v}$ are embedded accordingly:
\[
  \annotate{\mathcal{I}^*}{v} \gets \left\{ (\vec{J}, \vec{K})^\T \mid P\vec{K} + \vec{J} \in \annotate{\mathcal{I}}{v} \right\}, \text{ where } P=\diag(p_1, \ldots, p_n).
\]

\begin{example}
Tiling $\mathcal{I}=\{ 0\leq i< N \}$ into $t$ tiles, each of width $p$, yields
\[
  \mathcal{I}^* = \left\{ \vec{I}^* = (j, k)^\T \mid 0\leq j<p, 0\leq k<t=\left\lceil N/p \right\rceil \right\}.
\]
Embedding, for example, $\annotate{\mathcal{I}}{v_4}=\{ i < N \}$ into $\mathcal{I}^*$ yields $\annotate{\mathcal{I}^*}{v_4}\gets\{ (j, k)^\T \mid pk+j < N \}$.
\end{example}

After tiling, each tile $\vec{K}\in\mathcal{K}$ is assigned to a \ac{PE} of the \ac{TCPA} by the space mapping\footnote{This tiling and PE assignment is also known as \keyword{locally sequential, globally parallel} (LSGP) in the literature~\cite{nelis1988}.}:
\[
  \vec{\mathit{pe}}(\vec{K})\colon \mathcal{K}\mapsto \mathbb{N}^2 := \Phi\cdot \vec{K},\quad \Phi\in\mathbb{Z}^{2\times n},
\]
where $\Phi$ is the \keyword{allocation matrix}, which is chosen such that
$\vec{\mathit{pe}}(\vec{K}) = (k_r, k_c)^\T$, $1\leq r, c\leq n$ with $r\neq c$.
%that projects one dimension $r$ of $\vec{K}$ onto rows of \acp{PE} $\mathit{pe}_1$ and one dimension $c$ of $\vec{K}$ onto columns of \acp{PE} $\mathit{pe}_2$.
In the sequel, we use $\vec{K}$ and $\vec{\mathit{pe}}(\vec{K})$ interchangeably because in $\vec{K}$ all elements other than $k_r$ and $k_c$ are 0 by allowing only tile counts $t_r$ and $t_c$ to differ from 1.\footnote{For a one-dimensional mapping, either $k_r$ or $k_c$ is also 0.}

The distribution of iterations across multiple \acp{PE} may entail inter-processor communication for loop-carried dependences ($\vec{d}\neq 0$, for example $x[i]=y[i-1]$).
Here, we assume that communication always takes place between neighboring \acp{PE} (including the diagonal neighbors), that is dependence vectors never ``skip'' a tile;
in other words, we assume $p_i\geq d_i \forall 1\leq i \leq n$.
Which \acp{PE} do communicate for such a dependence $\vec{d}\neq 0$?
Suppose tiling maps an iteration $\vec{I}$ to tile $\vec{K}$.
Then the source iteration $\vec{I}-\vec{d}$ is on tiles $\{ \vec{K}-\vec{\theta} \}$ where $\vec{\theta}$ is from the set of tile displacements
\[
  \Theta(\vec{d}) := \left\{ \vec{\theta}=(\theta_1, \theta_2, \dots, \theta_n)^\T \mid \theta_i\in\{ 0, \operatorname{sign}(d_i) \} \right\}.
\]
The set of processor displacements---which processors $\{ \Phi\vec{K}-\vec{\delta} \}$ needs to communicate the result---is
\[
  \Delta(\vec{d}) := \left\{ \vec{\delta} \mid \vec{\delta}=\Phi\vec{\theta}\colon \forall \vec{\theta}\in\Theta(\vec{d}) \right\}.
\]

\begin{example}
Only $\vec{d}[e_4]=(1)$ is loop-carried (corresponding to equation $x[i]=y[i-1]$), resulting in the set of tile and processor displacements
\[
  \Theta(\vec{d}[e_4])=\{ (0), (1) \} \implies \Delta(\vec{d}[e_4]) := \left\{ \vec{\delta}_1=(0), \vec{\delta}_2=(1) \right\}.
\]
Consequently, the dependence results in communication from processor $k-1$ to processor $k$.
\end{example}

%% Modulo scheduling %%%%%%%%%%%%%%%%%%%%%%%%%%%%%%%%%%%%%%%%%%%%%%%%%%%%%%%%%%%%%%%%%%%%%%%%%%%%%%

\subsection{Scheduling of operations}
\label{sec:scheduling}

Next, \keyword{modulo scheduling}~\cite{rau1981}, a software pipelining technique, is performed to obtain a schedule that specifies a) when to start each \ac{PE}, b) when to start each iteration, c) when to start each operation within an iteration.
We assume that tiles are executed in parallel (since each is assigned to a different PE) in a wavefront-like fashion (to not violate data dependences), but iterations within a tile are executed sequentially (since \acp{PE} execute sequential programs).

Modulo scheduling constructs a cyclic schedule with period $\pi$, called the \keyword{initiation interval}, consisting of a linear part $\vec{\lambda}^*=(\vec{\lambda}^J, \vec{\lambda}^K)$ and a start offset $\tau[v]$ for each operation node $v$.
Given an iteration $\vec{I}^*$, the start time of $v$ then is
\[
  t_v(\vec{I}^*) = \vec{\lambda}^*\cdot \vec{I}^* + \annotate{\tau}{v},
\]
that is, $\vec{\lambda}^K$ determines the start times of the mapped \acp{PE}, $\vec{\lambda}^J$ determines the start times of the iterations assigned to a \ac{PE}, and $\annotate{\tau}{v}$ determines the relative start time of the operation.
Modulo scheduling also allocates a functional unit $\mathit{fu}[v]$ to execute operation $\annotate{op}{v}$ and selects the corresponding instruction $\mathit{mnemo}[v]$, including its latency $\annotate{l}{v}$.
Note that because we do not know the number and size of tiles in advance, we use \keyword{symbolic modulo scheduling} as introduced in \cite{witterauf2016}.

\begin{example}
For the sake of illustration, we assume each \ac{PE} has only two functional units \texttt{alu0} and \texttt{alu1}, both generic ALUs.
Because there are three different operations, we have to settle for an initiation interval $\pi=2$.
A linear schedule not violating the loop-carried dependence $\vec{d}[e_4]$ is $\vec{\lambda}^* = (\pi, \pi\cdot p)$.
For each operation node, the start offset $\tau$, allocated functional unit $\mathit{fu}$, and selected instruction $\mathit{mnemo}$ are shown in Figure~\ref{fig:rdg}.
We assume latency $\annotate{l}{v}=1\ \forall v$.
\end{example}

%%%%%%%%%%%%%%%%%%%%%%%%%%%%%%%%%%%%%%%%%%%%%%%%%%%%%%%%%%%%%%%%%%%%%%%%%%%%%%%%%%%%%%%%%%%%%%%%%%%

\subsection{Register allocation}
\label{sec:register-allocation}

Each edge in \iac{RDG} represents a dependence between two nodes:
the sink node consumes the value produced by the source node.
On \acp{TCPA}, these intermediate values are stored and communicated using different types of registers allocated per edge.
For convenience, in the following we simply write ``value of edge $e$'' instead of ``value produced/consumed via $e$'' or similar.
Two aspects determine the type of register that is allocated for an edge:
how long a value is \keyword{alive}, and whether it requires communication.
Inter-processor communication is required if an annotated dependence vector $\annotate{\vec{d}}{e}$ results in at least one processor displacement $\vec{\delta}\neq\vec{0}$.
To make this more explicit in the \ac{RDG}, we split each edge $e$ with $\vec{d}[e]\neq\vec{0}$ into $|\Delta(\vec{d}[e])|$ new edges $e^i$, all with the same annotations as $e$, but each additionally annotated with one of the processor displacements $\annotate{\vec{\delta}}{e^i}\gets\vec{\delta}_i\in\Delta(\annotate{\vec{d}}{e})$.
The original edge $e$ is removed.
\begin{example}
Edge $e_4$ corresponds to $|\Delta(\vec{d}[e_4])|=2$ processor displacements $\vec{\delta}_1=(0)$ and $\vec{\delta}_2=(1)$.
Therefore, it is split into two new edges: $e_4^1$ with $\annotate{\vec{\delta}}{e_4^1}\gets(0)$ and $e_4^2$ with $\annotate{\vec{\delta}}{e_4^2}\gets(1)$ as illustrated in Figure~\ref{fig:rdg}.
\end{example}

After splitting, for each edge $e=(v, w)\in E$, one or more registers are allocated according to the following classification.
If $e$ is a dependence edge, that is if both $v$ and $w$ are operation nodes \cite{hannig2009}:
\begin{itemize}
  \item If $\vec{d}[e]=\vec{0}$, the edge value is alive for $l'=\annotate{\tau}{w}-\annotate{\tau}{v}-\annotate{l}{v}$ time steps within a single iteration and a $k$-tuple of general-purpose registers $\annotate{\mathit{regs}}{e}\gets(r_1, r_2, \ldots, r_k)$ with $r_i\in R_\mathit{rd}$ is allocated, where $k=\lceil l' / \pi \rceil$.  
    Multiple registers are necessary if the edge value is still alive when the next value is produced, that is if $l'>\pi$.

  \item If $\vec{d}[e]\neq\vec{0}$ and $\vec{\delta}[e]=\vec{0}$, the edge value is alive for $\vec{\lambda}^J\vec{d}$ iterations, but only within the current \ac{PE}.
    A feedback register $\annotate{\mathit{reg}}{e}\gets r\in R_\mathit{fd}$ with depth $\vec{\lambda}^J\vec{d}$ is allocated.

  \item If $\vec{d}[e]\neq\vec{0}$ and $\vec{\delta}[e]\neq \vec{0}$, the edge value is alive across \acp{PE} and requires communication.
    Two registers are allocated:
    an input register $\annotate{\mathit{reg}^\mathit{read}}{e} \gets r \in R_\mathit{id}$ for reading on \ac{PE} $\Phi\vec{K}$ and an output register $\annotate{\mathit{reg}^\mathit{write}}{e} \gets r\in R_\mathit{od}$ for writing on the \ac{PE} $\Phi\vec{K}-\vec{\delta}[e]$.
    Additionally, a route from $\Phi\vec{K}-\vec{\delta}[e]$ to $\Phi\vec{K}$ on the interconnect network is allocated (see Section~\ref{sec:routing}).
\end{itemize}
Otherwise, if $e$ is an input edge, an input register $\annotate{\mathit{reg}}{e}\gets r\in R_\mathit{id}$ is allocated, and if $e$ is an output edge, an output register $\annotate{\mathit{reg}}{e}\gets r\in R_\mathit{od}$ is allocated;
in both cases, additionally an I/O access mapping is generated (Section~\ref{sec:symbolic-compilation-summary}).
For constant edges, no register is allocated (constants are immediate operands and do not require registers).

The conventional approach to register allocation---vertex coloring of an interference graph ~\cite{chaitin1981}---applies here, for which several optimal and heuristic solving methods exist.
%An interference graph is a graph that has a node for each entity that requires a register and an edge between any two nodes whose entities \keyword{interfere} with each other;
%that is, they cannot share the same register.
%Finding a $k$ vertex coloring of the interference graph then corresponds to an allocation of $k$ registers.
%Depending on the register type, \keyword{interference} is defined differently, as explained in more detail in the following sections.

%%%%%%%%%%%%%%%%%%%%%%%%%%%%%%%%%%%%%%%%%%%%%%%%%%%%%%%%%%%%%%%%%%%%%%%%%%%%%%%%%%%%%%%%%%%%%%%%%%%

\subsection{Routing of propagation channels}
\label{sec:routing}

The communication induced by each edge $e$ with a processor displacement $\vec{\delta}[e] \neq \vec{0}$ requires an interconnect route from an output port of \ac{PE} $\Phi\vec{K}$ to an input port of \ac{PE} $\Phi\vec{K}+\annotate{\vec{\delta}}{e}$ for each \ac{PE} $\vec{K}$ that produces at least one value of $e$.
However, because the tiling is performed symbolically, we do not know which \acp{PE} will be involved;
we therefore assume that the interconnect network is homogeneous---all interconnect wrappers have the same ports and adjacencies---, and utilizing this homogeneity, for each affected edge only one template route is allocated to be replicated across all \acp{PE} pairs during instantiation (see Section~\ref{sec:instantiation}).
Since this route forms a dedicated communication channel between equidistant pairs of \acp{PE} in order to propagate data across the \ac{TCPA}, we call it a \keyword{propagation channel}.
The set $R$ of propagation channels is found using a reduced topology graph.
\begin{definition}
A \keyword{reduced topology graph} $T$ is a directed graph that represents the topology of a homogeneous interconnect network.
The graph contains a node $v$ for each port in the interconnect wrapper and an edge for each possible connection between two ports weighted with their inter-processor distance.
In particular, there is an edge weighted $(0\ 0)^\T$ from each source port to each of its adjacent sink ports, as well as an edge from each wrapper sink port $\sinkport{\mathit{location}}{i}$ to its sibling port weighted by $\mathit{location}$: $(1\ 0)^\T$ for $\mathit{east}$, $(0\ 1)^\T$ for $\mathit{south}$, $(-1\ 0)^\T$ for $\mathit{west}$, and $(0\ -1)^\T$ for $\mathit{north}$.
\end{definition}

For convenience, assume $T$ has two polar nodes: $\mathit{Source}$, connected to all \ac{PE} output port nodes, and $\mathit{Sink}$, connected to all \ac{PE} input port nodes.
A propagation channel $\rho$ for \iac{RDG} edge $e$ is a path $(v_1, v_2, \ldots, v_{|\rho|})$ on $T$ from $\mathit{Source}$ to $\mathit{Sink}$ where the sum of weights equals $\annotate{\vec{\delta}}{e}$.
Routing $k=|\{ e\in E\colon \annotate{\vec{\delta}}{e}\neq 0 \}|$ propagation channels is thus equivalent to solving the \textit{$k$ node-disjoint exact-length paths problem}, with the relaxation that the paths of two propagation channels $\rho_1$ and $\rho_2$ may share the first $1 \leq r \leq \min(|\rho_1|, |\rho_2|)$ nodes if the two corresponding \ac{RDG} edges never have values alive simultaneously.
Then, the shared nodes (ports) of the reduced topology graph are never occupied at the same time.
However, sharing any node (port) but not its predecessors would imply that the port has two connected sources, which is not allowed.

The resulting routes and registers---the first node in each $\rho$ corresponds to a \ac{PE} output register, the last node to a \ac{PE} input register---are annotated to the corresponding edges in the \ac{RDG}.

\begin{figure}
  \centering
  \input{figures/tikz/channel-graph}
  \caption{%
    An example reduced topology graph, where each node represents an interconnect wrapper port, and each edge represents a possible connection (weights only shown if not $\vec{0}$).
    Visualized is a solution to routing two propagation channels, one for processor displacement $\vec{\delta}=(0\ 1)^\T$, depicted in blue, and one for processor displacement $\vec{\delta}=(1\ 1)^\T$, depicted in red.
    Assuming the corresponding \ac{RDG} edges never have values alive simultaneously, the first three nodes \sourceport{pe}{0}, \sinkport{east}{0}, and \sourceport{west}{0} can be shared.
    Node \sinkport{pe}{1} cannot be shared because then the port would have two different input connections.
  }
  \label{fig:reduced-topology-graph}
\end{figure}
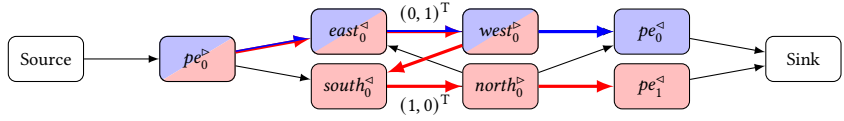

\begin{example}
Only edge $e_4^2$ has $\annotate{\vec{\delta}}{e_4^2} \neq \vec{0}$ and is therefore the only edge requiring a propagation channel; for example, $\annotate{\rho}{e_4^2} \gets (\sourceport{pe}{0}, \sinkport{east}{0}, \sourceport{west}{0}, \sinkport{pe}{0})$.
Consequently, the edge gets assigned registers $\annotate{\mathit{reg}^\mathit{write}}{e_4^2}\gets\texttt{od0}$ (from \sourceport{pe}{0}) and $\annotate{\mathit{reg}^\mathit{read}}{e_4^2}\gets\texttt{id0}$ (from \sinkport{pe}{0}).
For a more complex example, refer to Figure~\ref{fig:reduced-topology-graph}.
\end{example}

After tiling, scheduling, register allocation, and propagation channel routing, all necessary information is annotated to the \ac{RDG} to generate a polyhedral syntax tree, a symbolic representation of the set of programs that can be generated from the \ac{RDG} for any valid values of the parameters.

%%%%%%%%%%%%%%%%%%%%%%%%%%%%%%%%%%%%%%%%%%%%%%%%%%%%%%%%%%%%%%%%%%%%%%%%%%%%%%%%%%%%%%%%%%%%%%%%%%%

\subsection{Generation of a polyhedral syntax tree}
\label{sec:polyhedral-syntax-tree}

As motivated in the introduction, the generation of concrete \ac{PE} programs depends both on the loop bounds and the number of allocated \acp{PE}.
However, from the annotated \ac{RDG}, all instructions, their condition spaces, functional unit, and time offset have meanwhile be determined at compile time; only their composition depends on the still unknown parameter values.
A \keyword{polyhedral syntax tree}~\cite{witterauf2019} represents this information hierarchically such that it is equivalent to the forest of \ac{PE} program syntax trees over all parameter values.
Its building blocks are so-called \keyword{fragments}.

\begin{definition}[\cite{witterauf2019}]
A \keyword{fragment} $F$ is any syntactic constituent of a program.
\end{definition}

For example, the assembly instruction \texttt{addi\,rd0\,rd1\,10} can be structured into five fragments: the mnemonic \texttt{addi}, the registers \texttt{rd0} and \texttt{rd1}, the literal \texttt{10}, and finally the entire instruction itself.
Whether an operation is executed within an iteration $\vec{I}^*$, for example, depends on its condition space.
Fragments may thus be iteration-dependent.

\begin{definition}[\cite{witterauf2019}]
A \keyword{polyhedral fragment} $f(\vec{I})$ maps an iteration $\vec{I}\in\mathcal I$ to a fragment $F$.
\end{definition}

We denote specific polyhedral fragments by \keyword{polyhedral $\langle$fragment name$\rangle$}, for example polyhedral register or polyhedral instruction.
Since fragments are syntactic, representation as a tree is natural.

\begin{definition}[Adapted from \cite{witterauf2019}]\label{def:polyhedral-fragment-tree}
A \keyword{polyhedral syntax tree (PST)} is a triple $f=(\mathcal{I}, a, G)$ of a condition space $\mathcal{I}$, a tuple $a$ of attributes, and a set of children $G$, each of which is again a polyhedral syntax tree.
To avoid ambiguity, we write $\domain(f)$ for $\mathcal{I}$, $\attrs(f)$ for $a$, and $\children(f)$ for $G$ of $f$, but we use the term node for both a polyhedral syntax tree itself and its children.
Each node is of one of two types:
If $\attrs(f)=(F)$, that is if the tuple only contains a fragment $F$, then $f$ is a \keyword{fragment node}.
Otherwise, $f$ is a \keyword{meta node} that stores implementation-specific syntactic meta-information.
A node of a polyhedral syntax tree satisfies the following properties regarding its immediate children:
All children are of the same type;
if the children are fragment nodes, their condition spaces must be disjoint;
if it has children, its condition space is the union of its children's condition spaces;
no two children may have the same attribute values.
The \keyword{evaluation} $f(\vec{I})$ of a polyhedral syntax tree is the sub-tree where all nodes $g$ with $\vec{I}\notin\domain(g)$ are removed.
\end{definition}

The evaluation of a PST at a concrete iteration $\vec{I}$ results in a syntax tree that represents the sequence of instructions \keyword{issued} in that iteration.
Thus, concatenating the instruction sequences in execution order for all $\vec{I}$ in the iteration space $\mathcal{I}$ yields an unrolled assembly program for the entire loop.
This observation serves as the basis for an efficient program generation algorithm in Section~\ref{sec:control-flow-analysis}.

\newcommand*{\fragmentdomain}[1]{$[\{ #1 \}]$}
\newcommand*{\metalabel}[1]{\textit{\textlangle #1\textrangle}}

\begin{example}
The PST generated from the RDG of the running example is shown below, showing the condition spaces (in brackets) after tiling only for the leaf nodes (since all other condition spaces are unions of these).
Fragment nodes are set in \texttt{typewriter}, meta nodes in \metalabel{italic}.
For \acp{TCPA}, a PST has the following semantics:
The second level represents the functional unit program $\mathit{fu}$, the third the time offset $\tau$, the fourth the instruction $\mathit{mnemo}$, and the last two the instruction's operands.
\begin{center}
  \small
  $f\coloneqq$
  \begin{forest}
    for tree={grow'=east, child anchor=west, anchor=west, s sep=0pt, tier/.pgfmath=level()}
    [\metalabel{root}, baseline, anchor=base,
      [\texttt{alu0}
        [\metalabel{$\tau$: 0}
          [\texttt{move}
            [\metalabel{dest}
              [$g_1\coloneqq$ \texttt{rd0} \fragmentdomain{pk + j < N}]
            ]
            [\metalabel{src}
              [$g_2\coloneqq$ \texttt{id0} \fragmentdomain{j = 0}]
              [$g_3\coloneqq$ \texttt{rd1} \fragmentdomain{j > 0 \wedge pk+j<N}]
            ]
          ]
        ]
      ]
      [\texttt{alu1}
        [\metalabel{$\tau$: 1}
          [\texttt{shri}
            [\metalabel{dest}
              [$g_4\coloneqq$ \texttt{od0} \fragmentdomain{j = p - 1 \wedge k < t - 1}]
              [$g_5\coloneqq$ \texttt{rd1} \fragmentdomain{j < p - 1 \wedge pk+j<N}]
            ]
            [\metalabel{srcA}
              [$g_6\coloneqq$ \texttt{rd0} \fragmentdomain{pk+j<N}]
            ]
            [\metalabel{srcB}
              [$g_7\coloneqq$ \texttt{1} \fragmentdomain{pk+j<N}]
            ]
          ]
        ]
        [\metalabel{$\tau$: 2}
          [\texttt{andi}
            [\metalabel{dest}
              [$g_8\coloneqq$ \texttt{od1} \fragmentdomain{pk+j<N}]
            ]
            [\metalabel{srcA}
              [$g_9\coloneqq$ \texttt{rd0} \fragmentdomain{pk+j<N}]
            ]
            [\metalabel{srcB}
              [$g_{10}\coloneqq$ \texttt{1} \fragmentdomain{pk+j<N}]
            ]
          ]
        ]
      ]
    ]
  \end{forest}
\end{center}
In the following, we refer to elements of $\attrs(f)$ using annotation syntax, for example $\annotate{F}{g_1}=\texttt{rd0}$.
The generation of a PST from \iac{RDG} is described in detail in previous work~\cite{witterauf2019}.
\end{example}

%%%%%%%%%%%%%%%%%%%%%%%%%%%%%%%%%%%%%%%%%%%%%%%%%%%%%%%%%%%%%%%%%%%%%%%%%%%%%%%%%%%%%%%%%%%%%%%%%%%

\subsection{Compilation of access mappings}

Still missing is information about accesses to external data, that is, to the input and output variables.
To represent this information, for each input and output edge, an \keyword{access mapping} $a$ is compiled from the edge's annotations.
\begin{definition}
An access mapping is a four-tuple $a=\left(\mathit{reg}, x, \alpha, \mathcal{A} \right)$ that maps all accesses to register $\mathit{reg}$ in iterations $\vec{I}\in\mathcal{A}$ to variable element $x[\alpha(\vec{I})]$.
\end{definition}
Given an input or output edge $e$, the indexing function is $\alpha = (\annotate{Q}{e}, \annotate{\vec{d}}{e})$, its access space $\mathcal{A}=\annotate{\mathcal{I}}{v}$ (input access) or $\mathcal{A}=\annotate{\mathcal{I}}{w}$ (output access), the register $\mathit{reg}=\annotate{\mathit{reg}^\mathit{read}}{e}$ (input access) or $\mathit{reg}=\annotate{\mathit{reg}^\mathit{write}}{e}$ (output access), and the associated input variable $x=\annotate{x}{v}$ or output variable $x=\annotate{x}{w}$.
We denote the set of all access mappings in a symbolic configuration $A$.

\begin{example}
The access mappings $A=\{ a_\mathit{in}, a_\mathit{bits} \}$ are as follows:
\begin{itemize}
  \item $a_\mathit{in}$: Edge $e_1$ represents accesses to the input scalar $\mathit{in}$.
  We model scalars as zero-dimensional variables, that means, the indexing function is $\alpha=(\annotate{Q}{e_1}=()\in\mathbb{Z}^{0\times n}, \annotate{\vec{d}}{e}=()\in\mathbb{Z}^0)$. (For these empty matrices, we assume that $\mathbb{Z}^{0\times n}\cdot\mathbb{Z}^{n} \in \mathbb{Z}^{0}$.)
    Its access domain is $\mathcal{A}=\annotate{\mathcal{I}}{v_1}=\{ i = 0 \}$, the allocated input register $\mathit{reg}=\texttt{id0}$.
  \item $a_\mathit{bits}$: Edge $e_7$ represents accesses to the one-dimensional output variable $\mathit{bits}$.
    The indexing function is $\alpha=(\annotate{Q}{e_7}=(1), \annotate{\vec{d}}{e_7}=(0))$.
    It is written to in each iteration, meaning $\mathcal{A}=\annotate{\mathcal{I}}{v_7}=\{ i < N \}$.
    The allocated output register is $\mathit{reg}=\texttt{od0}$.\qedhere
\end{itemize}
\end{example}

%%%%%%%%%%%%%%%%%%%%%%%%%%%%%%%%%%%%%%%%%%%%%%%%%%%%%%%%%%%%%%%%%%%%%%%%%%%%%%%%%%%%%%%%%%%%%%%%%%%

\subsection{Summary: symbolic compilation}
\label{sec:symbolic-compilation-summary}

After these steps, a symbolic configuration consisting of the following parts is obtained:
A polyhedral syntax tree $f$ to generate \ac{PE} programs from, a symbolic schedule $\vec{\lambda}^*$ that schedules tiles (\acp{PE}) in parallel, iterations within a tile sequentially, a set $R$ of propagation routes that will be replicated across the allocated \ac{TCPA} region, and a set $A$ of access mappings from which the I/O buffer and address generator configurations will be generated.

%% file: figures/tikz/channel-graph.tex
\begin{tikzpicture}[
  font=\footnotesize,
  graph node/.style={
    draw,
    rounded corners=3pt,
    minimum width=1cm,
    minimum height=0.6cm,
    inner sep=0pt,
    align=center,
  },
  node distance=0.25cm and 1cm,
  route a/.style={fill=red!25},
  route b/.style={fill=blue!25},
  route a path/.style={->, >=latex, very thick, red},
  route b path/.style={->, >=latex, very thick, blue},
  tight background,
]

  \node[graph node] (pe source) {Source};
  \node[graph node, right=of pe source] (p-o-0) {};
  \node[graph node, route a, above right=of p-o-0, yshift=-0.5cm] (e-o-0) {};
  \node[graph node, route a, below right=of p-o-0, yshift=0.5cm] (s-o-0) {\sinkport{south}{0}};
  \node[graph node, route a, right=of e-o-0] (w-i-0) {};
  \node[graph node, route a, right=of s-o-0] (n-i-0) {\sourceport{north}{0}};
  \node[graph node, route b, right=of w-i-0] (p-i-0) {\sinkport{pe}{0}};
  \node[graph node, route a, right=of n-i-0] (p-i-1) {\sinkport{pe}{1}};
  \node[graph node, above right=of p-i-1, yshift=-0.5cm] (pe sink) {Sink};

  \fill[route b] (p-o-0.south west) [rounded corners=3pt] -- (p-o-0.north west) [rounded corners=3pt] -- (p-o-0.north east) -- cycle;
  \fill[route a] (p-o-0.south west) [rounded corners=3pt] -- (p-o-0.south east) [rounded corners=3pt] -- (p-o-0.north east) -- cycle;
  \node[graph node] at (p-o-0) {\sourceport{pe}{0}};

  \fill[route b] (e-o-0.south west) [rounded corners=3pt] -- (e-o-0.north west) [rounded corners=3pt] -- (e-o-0.north east) -- cycle;
  \fill[route a] (e-o-0.south west) [rounded corners=3pt] -- (e-o-0.south east) [rounded corners=3pt] -- (e-o-0.north east) -- cycle;
  \node[graph node] at (e-o-0) {\sinkport{east}{0}};

  \fill[route b] (w-i-0.south west) [rounded corners=3pt] -- (w-i-0.north west) [rounded corners=3pt] -- (w-i-0.north east) -- cycle;
  \fill[route a] (w-i-0.south west) [rounded corners=3pt] -- (w-i-0.south east) [rounded corners=3pt] -- (w-i-0.north east) -- cycle;
  \node[graph node] at (w-i-0) {\sourceport{west}{0}};

  %\path (p-o-0) -- (e-o-0) coordinate[at start](h1) coordinate[at end] (h2);
  %\draw[->]($(h1)!4pt!90:(h2)$)-- node [auto=left] {} ($(h2)!4pt!-90:(h1)$); 
  %\draw[->]($(h1)!4pt!-90:(h2)$)-- node [auto=right] {} ($(h2)!4pt!90:(h1)$);

  \path
    (e-o-0) edge node[above] {$(0, 1)^\T$} (w-i-0)
    (s-o-0) edge node[below] {$(1, 0)^\T$} (n-i-0);

  \path[->, >=latex]
    (pe source) edge (p-o-0)
    (p-o-0) %edge (e-o-0)
            edge (s-o-0)
    (w-i-0) edge (p-i-0)
    (n-i-0) edge (p-i-0)
            edge (e-o-0)
    (p-i-0) edge (pe sink)
    (p-i-1) edge (pe sink)
    ;
  \path[route a path]
    (p-o-0) edge (e-o-0)
    (e-o-0) edge (w-i-0)
    (w-i-0) edge (s-o-0)
    (s-o-0) edge (n-i-0)
    (n-i-0) edge (p-i-1);
    ;
  \path[route b path]
    (w-i-0) edge (p-i-0);

  \begin{scope}
    \path (p-o-0) -- (e-o-0) coordinate[at start] (h1) coordinate[at end] (h2);
    \clip (h1) -- ($(h1)!4pt!90:(h2)$) -- ($(h2)!4pt!-90:(h1)$) -- (h2) -- cycle; 
    \draw[route b path] (h1) -- (h2);
  \end{scope}
  \begin{scope}
    \path (e-o-0) -- (w-i-0) coordinate[at start] (h1) coordinate[at end] (h2);
    \clip (h1) -- ($(h1)!4pt!90:(h2)$) -- ($(h2)!4pt!-90:(h1)$) -- (h2) -- cycle; 
    \draw[route b path] (h1) -- (h2);
  \end{scope}

\end{tikzpicture}

%% file: sections/runtime.tex
\section{Instantiation}
\label{sec:instantiation}

Instantiation is the generation of a \keyword{concrete configuration} from a given \keyword{symbolic configuration} and an assignment of values to the parameters---the concrete loop bounds, the number of allocated processing elements, and the memory layouts of input and output arrays.
It comprises these steps:
\begin{enumerate}
  \item \textbf{Concretization} substitutes all occurrences of parameters in the symbolic configuration, for example in the iteration space $\mathcal{I}^*$, with their assigned values.
    Using the concretized schedule, the feedback register depths ($\vec{\lambda}^J\vec{d}/\pi$) are computed.
  
  \item \textbf{Program instantiation} is the most complex step and further divided into three sub-steps:
    \begin{enumerate}      
      \item \textbf{Control flow analysis} (Section~\ref{sec:control-flow-analysis}) first determines a set of processor classes, that is, a partitioning of $\mathcal{K}$ into subsets $\mathcal{P}_\mathit{pc}$ of \acp{PE} that will execute the same \ac{PE} program.
        For each processor class $\mathcal{P}_\mathit{pc}$ and each functional unit $\mathit{fu}$, a control flow graph $\mathit{CFG}_{\mathit{pc}, \mathit{fu}}$ is generated from the specialized polyhedral syntax tree $f_\mathit{pc}=f(\mathcal{P}_\mathit{pc}\oplus\mathcal{J})$ and intra-tile schedule $\vec{\lambda}^J$.
        Each node in $\mathit{CFG}_{\mathit{pc}, \mathit{fu}}$ represents an atomic sequence of instructions and each edge represents a branch annotated with its \keyword{transition space} $\mathcal{T}$ (the iterations in which the branch is taken).
      \item \textbf{Control signal allocation} (Section~\ref{sec:control-signal-allocation}) allocates a set of binary control signals that, for each iteration $\vec{J}\in\mathcal{J}$, encode for all transition spaces across all processor classes and control flow graphs whether $\vec{J}\in\mathcal{T}$.
        The corresponding control signal values are annotated back to the control flow graph edges and used later for generating branch conditions.
      \item \textbf{Program generation} (Section~\ref{sec:program-generation}) generates a \ac{PE} program for each processor class $\mathcal{P}_\mathit{pc}$ that contains a functional unit program for each $\mathit{CFG}_{\mathit{pc}, \mathit{fu}}$:
      The instruction sequences represented by the nodes are concatenated and branch instructions generated according to the control signal values annotated to the edges.
    \end{enumerate}

  \item \textbf{Periphery instantiation} generates a concrete configuration for the global controller from the allocated set of control signals (therefore also described in Section~\ref{sec:control-signal-allocation}).
    Furthermore, for each access mapping $a\in A$, each involved \ac{PE} is connected to a memory bank, whose address generator configuration is generated according to the indexing function $\alpha$ (Section~\ref{sec:io-access-generation}).

  \item \textbf{Interconnect instantiation} replicates the propagation channels for all allocated interconnect wrappers and incorporates any additional routes from the I/O routing, resulting in the concrete interconnect configuration (see Figure~\ref{fig:config-overview} for the running example).
\end{enumerate}

Instantiation is, in general, performed at runtime.
%Runtime instantiation allows the system to dynamically allocate regions of a \ac{TCPA}, for example to support resource-aware programming, such as in Invasive Computing~\cite{teich2011}.
Making runtime instantiation viable requires the above steps to be efficient---that is, to have low-degree polynomial time and space complexity---and scale well with an increasing number of \ac{PE}s.
The latter especially matters for the most complex part of instantiation: program instantiation.

\begin{example}
Suppose we choose as loop bound $N=16$ (extracting the first 16 bits from $\mathit{in}$) and allocate $t=3$ \acp{PE}, resulting in a minial tile size $p=\lceil N/t \rceil=6$, which is an imperfect tiling (tile $k=2$ is not full since $3\cdot 6=18>N$).
Concretization yields the schedule $\vec{\lambda}=(12, 2)$.
%, resulting in the \ac{PE} start times $t(0)=0$, $t(1)=12$, and $t(2)=24$.
The allocated feedback register \texttt{fd0} has a depth of $\vec{\lambda}^J\annotate{\vec{d}}{e_4^0} / 2 = 1$.
\end{example}

%%%%%%%%%%%%%%%%%%%%%%%%%%%%%%%%%%%%%%%%%%%%%%%%%%%%%%%%%%%%%%%%%%%%%%%%%%%%%%%%%%%%%%%%%%%%%%%%%%%

\subsection{Control flow analysis}
\label{sec:control-flow-analysis}

Generating a program for each \ac{PE} separately does not scale to arbitrary \ac{TCPA} sizes;
however, due to the regularity of loops, large subsets of \acp{PE} execute the same program.
Thus, by generating each distinct \ac{PE} program only once, program instantiation scales well because the number of distinct programs across \acp{PE} is bounded even if the number of \acp{PE} keeps increasing.
But is it possible to determine whether multiple \acp{PE} will be configured with the same program without actually generating their programs?
Yes, by using the information in the polyhedral syntax tree (PST):
We consider the programs of two \acp{PE} $\vec{K}_1$ and $\vec{K}_2$ equal if \keyword{specialization} yields the same PST for both.
\begin{definition}
  Given a condition space after tiling $\mathcal{I}^*\subseteq\mathcal{J}\oplus\mathcal{K}$, the function
  \[
    \operatorname{split}\colon \vec{K}, {\mathcal{I}^*} \mapsto \widehat{\mathcal{J}} = \{ \vec{J} \in \mathcal{J} \mid (\vec{J}, \vec{K})^T \in \mathcal{I}^* \}
  \]
  maps a tile $\vec{K}\in\mathcal{K}$ to its \keyword{intra-tile domain} $\widehat{\mathcal{J}}$ within $\mathcal{I}^*$, that is the set of iterations $\vec{J}$ within tile $\vec{K}$ that lie in $\mathcal{I^*}$.
\end{definition}
\begin{definition}\label{def:specialization}
  Given a polyhedral syntax tree $f$ with condition space after tiling $\mathcal{I}^*=\domain(f)\subseteq \mathcal{J}\oplus\mathcal{K}$, \keyword{specialization} for a tile $\vec{K}\in\mathcal{K}$, denoted $f \triangleright \vec{K}$, recursively maps $\mathcal{I}^*$ to the intra-tile domain of $\vec{K}$ within $\mathcal{I}^*$:
  \[
    f \triangleright \vec{K} \coloneqq \left(\operatorname{split}(\vec{K}, \domain(f)), \attrs(f), \left\{ g \triangleright \vec{K} \mid g\in \operatorname{children}(f) \right\}\right)
  \]
\end{definition}
\begin{example}
Using $p=6$, $t=3$ and $N=16$, the condition space $\domain(g_1)=\{ pk+j<N \}$ concretizes to $\{ 6k + j < 16 \}$.
For tile $k=2$, the corresponding intra-tile domain (iterations $j$ that satisfy $6\cdot 2 + j < 16$) is $\{ 0\leq j\leq 3 \}$.
Consequently, the specialization $g_1\triangleright (2)$ yields $g_1\coloneqq$ \texttt{rd0} \fragmentdomain{0\leq j\leq 3}.
\end{example}
We use Definition~\ref{def:specialization} to partition the inter-tile space $\mathcal{K}$ into \keyword{processor classes} $\mathcal{P}_\mathit{pc}$, each of which is a set of \acp{PE} that result in the same specialized PST and thus program.
But when is $f\triangleright \vec{K}_1 = f\triangleright \vec{K}_2$ for two distinct \acp{PE} $\vec{K}_1$ and $\vec{K}_2$?
Since specialization only transforms condition spaces and all condition spaces in a PST are unions of its children's condition spaces, the two specializations are equal if $\domain(g\triangleright \vec{K}_1) = \domain(g\triangleright \vec{K}_2)$ for all leaves $g$ of $f$.
Hence, we must investigate when $\vec{K}_1$ and $\vec{K}_2$ result in the same intra-tile domain within a condition space $\mathcal{I}^*$.

\begin{definition}\label{def:intra-tile-patterns}
Given a condition space after tiling $\mathcal{I}^*\subseteq\mathcal{K}\oplus\mathcal{J}$, the set of tiles with the same intra-tile domain $\widehat{\mathcal{J}}$, called its \keyword{inter-tile domain} $\widehat{\mathcal{K}}$, is given by the function
\[
  \operatorname{tiles}\colon \widehat{\mathcal{J}}, \mathcal{I}^* \mapsto \widehat{\mathcal{K}} = \{ \vec{K} | \vec{K}\in\mathcal{K} \wedge \operatorname{split}(\vec{K}, \mathcal{I}^*) = \widehat{\mathcal{J}} \}.
\]
We call $\widehat{\mathcal{I}}=(\widehat{\mathcal{J}}, \widehat{\mathcal{K}})$ the \keyword{intra-tile pattern} of $\widehat{\mathcal{K}}$ within $\mathcal{I}^*$.
% and, as a short-hand, write $\widehat{\mathcal{I}}\colon\widehat{\mathcal{K}}\rightarrow \widehat{\mathcal{J}}$.
We denote the set of all intra-tile patterns of $\mathcal{I}^*$ as $\mathbb{I}_{\mathcal{I}^*}$, which always corresponds to a partitioning of $\mathcal{K}$.
\end{definition}

\begin{example}
For $\mathcal{I}^*=\domain(g_1)$, there are $|\mathbb{I}_{\mathcal{I}^*}|=2$ intra-tile patterns:
$\widehat{\mathcal{I}}_1=(\{ 0 \leq j \leq 5 \}, \{ k<2 \})$ (full tiles) and $\widehat{\mathcal{I}}_2=(\{ 0 \leq j \leq 3 \}, \{ k=2 \})$ (partial tile).
\end{example}

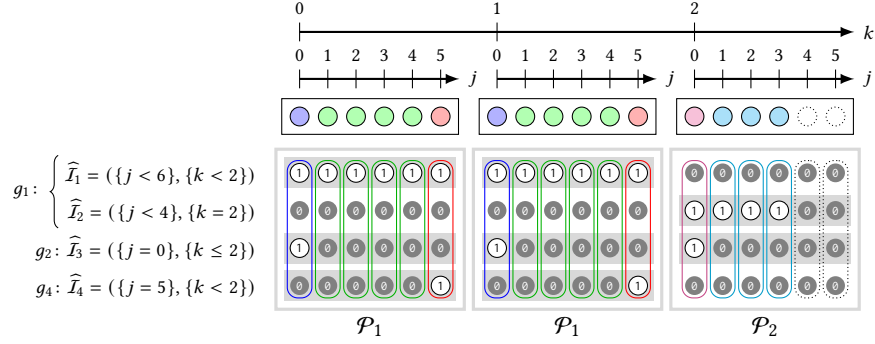
\begin{figure}
  \centering
  \input{figures/tikz/processor-classes.tex}%
  \vspace{-0.25cm}
  \caption{Illustration of the relation between intra-tile patterns and processor classes.
    The top shows the iteration space $\mathcal{I}^*$ of the running example after tiling with loop bounds $N=16$, tile size $p=6$, and tile count $t=3$.
    Each color represents a distinct sequence of instructions issued in that iteration $\vec{I}^*=(j, k)^\T$.
    Below, the intra-tile patterns of the condition spaces of the leaves $g_1$, $g_2$, and $g_4$ of the PST $f$ are visualized (omitting the others for brevity since they do not result in distinct intra-tile patterns).
    Each row visualizes one intra-tile pattern $\widehat{\mathcal{I}}_i=(\widehat{\mathcal{J}}_i, \widehat{\mathcal{K}}_i)$:
    There is a 1 if $(j)\in\widehat{\mathcal{J}}_i\wedge (k)\in\widehat{\mathcal{K}}_i$, a 0 otherwise, and a gray box in the background if $(k)\in\widehat{\mathcal{K}}_i$.
    Reading the column of an iteration $\vec{I}^*$ as a binary number yields an encoding for the evaluation $f(\vec{I}^*)$;
    for example, for the left-most iteration, the encoding is \texttt{1010}, meaning that $g_1$ and $g_2$ remain after evaluation, but $g_4$ is removed.
    Therefore, two iterations $\vec{I}^*_1$ and $\vec{I}^*_2$ with the same encoding---visualized by color in the figure---issue the same combination of instructions because $f(\vec{I}^*_1)=f(\vec{I}^*_2)$.
    If two \acp{PE} have the same coloring for all iterations within its assigned tile, they thus have the same program and belong to the same processor class.
    Having the same coloring is equivalent to being in the same set of inter-tile domains, that is, having the same pattern of gray boxes in the figure.
    Hence, the examples results in two processor classes $\mathcal{P}_1$ and $\mathcal{P}_2$.
  }
  \label{fig:processor-classes}
\end{figure}

Definition~\ref{def:intra-tile-patterns} implies that if $\vec{K}_1$ and $\vec{K}_2$ are part of the same intra-tile pattern $\widehat{\mathcal{I}}=(\widehat{\mathcal{J}}, \widehat{\mathcal{K}})$ of a leaf $g$'s condition space, that is if both $\vec{K}_1\in\widehat{\mathcal{K}}$ and $\vec{K}_2\in\widehat{\mathcal{K}}$, specialization maps them to the same intra-tile pattern $\widehat{\mathcal{J}}$, making $\domain(g\triangleright \vec{K}_1)=\domain(g\triangleright \vec{K}_2)$.
Consequently, the determination of processor classes depends only on the inter-tile domains:
If $\vec{K}_1$ and $\vec{K}_2$ are in the same inter-tile domain for each leaf node of $f$, they result in the same specialized PST.
Figure~\ref{fig:processor-classes} illustrates this relation between processor classes and intra-tile patterns, as well as why two \acp{PE} share the same program if the specialized PSTs are equal.
To formalize this visual intuition, let $\mathbb{K}\coloneqq\{ \widehat{\mathcal{K}}_1, \widehat{\mathcal{K}}_2, \ldots \}$ be the set of all inter-tile domains annotated to the leaves of $f$:
\[
  \mathbb{K} = \left\{ \widehat{\mathcal{K}} \mid \exists g\in\operatorname{leaves}(f)\colon \exists (\widehat{\mathcal{J}}, \widehat{\mathcal{K}}') \in \mathbb{I}_{\domain(g)}\colon \widehat{\mathcal{K}}=\widehat{\mathcal{K}}' \right\}.
\]
For each \ac{PE} $\vec{K}\in\mathcal{K}$, there is a partitioning of $\mathbb{K}$ into two subsets:
$\mathbb{K}^+_{\vec{K}}$, containing all $\widehat{\mathcal{K}}_i$ such that $\vec{K}\in\widehat{\mathcal{K}}_i$, and $\mathbb{K}^-_{\vec{K}}$, containing all $\widehat{\mathcal{K}}_i$ such that $\vec{K}\notin\widehat{\mathcal{K}}_i$.
As per the reasoning above, if the partitioning is equal for two \acp{PE} $\vec{K}_1$ and $\vec{K}_2$, they are part of the same inter-tile domain for each leaf node of $f$, meaning the specialized PSTs are equal as well.
Given $\vec{K}$, the set $\mathcal{P}_{\vec{K}}$ of \acp{PE} with the same partitioning is
\begin{equation}\label{eq:processor-class}
  \mathcal{P}_{\vec{K}} =
  \bigcap\limits_{\widehat{\mathcal{K}}\in\mathbb{K}^+_{\vec{K}}} \widehat{\mathcal{K}}
  \cap 
  \bigcap\limits_{\widehat{\mathcal{K}}\in\mathbb{K}^-_{\vec{K}}} \overline{\widehat{\mathcal{K}}},
\end{equation}
making the \keyword{set of processor classes} $\mathit{PC}=\{ \mathcal{P}_{\vec{K}} \mid \vec{K}\in\mathcal{K} \}$.
By rearranging Equation~\eqref{eq:processor-class} as in Algorithm~\ref{alg:processor-classes}, we can avoid a time complexity proportional to the number of \acp{PE} $|\mathcal{K}|$ to obtain $\mathit{PC}=\{ \mathcal{P}_1, \ldots \}$ as output $D$ using $\mathcal{D}=\mathcal{K}$ as initial partitioning and $C=\mathbb{K}$ as set of conditions.

\algnewcommand{\IfThen}[2]{% \IfThen{<if>}{<then>}
  \State \algorithmicif\ #1\ \algorithmicthen\ #2}

\begin{algorithm}
  \caption{Partition $\mathcal{D}$ into subsets $D=\{ \mathcal{D}_1, \ldots \}$ according to conditions $C=\{ \mathcal{C}_1, \ldots \}$}
  \label{alg:processor-classes}
  \begin{algorithmic}
    \State $D \gets \{ \mathcal{\mathcal{D}} \}$ \Comment{start with full space $\mathcal{D}$ as ``partitioning''}
    \For{$\mathcal{C} \in C$} \Comment{for each condition, partition current set of subsets}
      \State $\mathit{D}' \gets \emptyset$
      \For{$\mathcal{D}'\in\mathit{D}$} \Comment{partition each subset according to condition into up to two subsets}
        \IfThen{$\mathcal{D}'\cap\mathcal{C} \neq\emptyset$}%
          {$D' \gets D' \cup\{ \mathcal{D}'\cap \mathcal{C} \}$}
        \IfThen{$\mathcal{D}'\cap\overline{\mathcal{C}}\neq\emptyset$}
          {$D' \gets D' \cup\{ \mathcal{D}'\cap\overline{\mathcal{C}} \}$}
      \EndFor
      \State $\mathit{D} \gets D'$
    \EndFor
  \end{algorithmic}
\end{algorithm}

\begin{example}
Assuming the same tiling parameters as before, there are three distinct inter-tile domains: $\widehat{\mathcal{K}}_1=\{ k < 2 \}$, $\widehat{\mathcal{K}}_2=\{ k \leq 2 \}$, and $\widehat{\mathcal{K}}_3=\{ k = 2 \}$.
Consider \ac{PE} $k=0$:
$\mathbb{K}^+_K=\{ \widehat{\mathcal{K}}_1, \widehat{\mathcal{K}}_2 \}$ and $\mathbb{K}^-_K=\{ \widehat{\mathcal{K}}_3 \}$.
According to Equation~\eqref{eq:processor-class}, all \acp{PE} in the same processor class are then
\[
    \widehat{\mathcal{K}}_1 \cap \widehat{\mathcal{K}}_2 \cap \overline{\widehat{\mathcal{K}}_3} = \{ k < 2 \} \cap \{ k \leq 2 \} \cap \{ k \neq 2 \} \equiv \{ k < 2 \}.
\]
Overall, there are two processor classes:
$\mathit{PC} = \{ \mathcal{P}_1=\{ k < 2 \}, \mathcal{P}_2=\{ k = 2 \} \}$.
\end{example}

Next, for each processor class $\mathcal{P}_\mathit{pc}$, a control flow graph for each functional unit in the specialized PST $f_\mathit{pc}=f \triangleright \vec{K}$, $\vec{K}\in\mathcal{P}_\mathit{pc}$ is constructed.
However, generating compact programs necessitates exploiting repetition, but successive iterations may overlap in time due to software pipelining.
\begin{example}
  Suppose we fully unroll the functional unit program described by \texttt{alu1} of $f_1=f\triangleright (0)$ into pseudo-assembly (time offset $\tau$ in brackets):
  {\small
  \begin{verbatim}
    [0]             nop             // j = 0 starts (prolog)
    [1]             shri fd0 rd0 1  
    [2] [0]         andi od1 rd0 1  // j = 1 starts  --\
        [1]         shri fd0 rd0 1                     |
        // ...                                         |-- repetition
        [2] [0]     andi od1 rd0 1  // j = 4 starts    |
            [1]     shri fd0 rd0 1                   --/
            [2] [0] andi od1 rd0 1  // j = 5 starts
                [1] shri od0 rd0 1
                [2] andi od1 rd0 1  // epilog starts\end{verbatim}}
  Observe that iterations overlap and that between $1\leq j \leq 4$, the same two instructions repeat.
  A timing-equivalent \emph{compact} program is the following, arranged by iteration:
  {\small\begin{verbatim}
        nop            ; shri fd0 rd0 1                      // j = 0
    L1: andi od1 rd0 1 ; shri fd0 rd0 1 ; goto L1 if j <= 4  // 1 <= j <= 4
        andi od1 rd0 1 ; shri od0 rd0 1                      // j = 5        
        andi od1 rd0 1                                       // epilog\end{verbatim}}
  The $\pi=2$ instructions that repeat originate from different evaluations of $f_1((j))$---\texttt{andi} from iterations $0\leq j\leq 3$ and \texttt{shri} from iterations $1\leq j\leq 4$.
  How can we find such a compact program \emph{without} unrolling the program?
\end{example}

In a modulo-scheduled functional unit program, the next iteration is issued every $\pi$ time steps, making the execution of the program a sequence of assembly \keyword{kernels} of length $\pi$ \keyword{slots} each and each slot housing one instruction.
(In the compact program example above, each line is a kernel.)
Control flow therefore only changes each $\pi$ time steps---each kernel is executed atomically---, an observation we use to reformulate the problem:
How do we determine all kernels necessary for program instantiation and build a corresponding control flow graph?

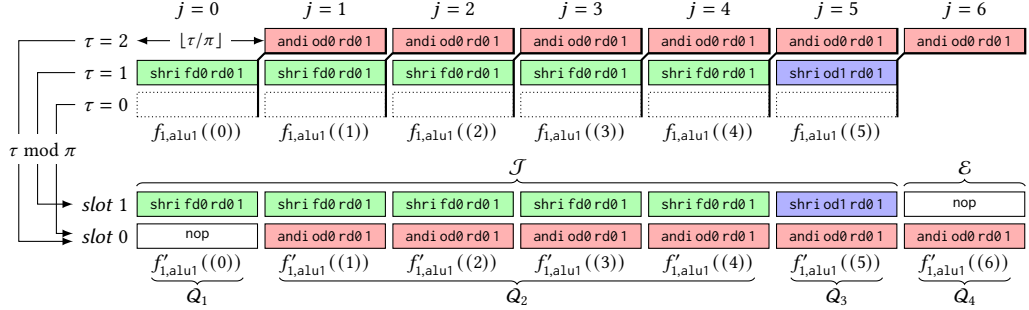
\begin{figure}
  \centering
  \input{figures/tikz/kernel-classes}%
  \vspace{-0.25cm}
  \caption{Formation of kernels in $f_{1, \texttt{alu1}}$ with $\pi=2$.
    Each column represents an iteration $0\leq j\leq 6$ ($j=6$ being a ``pseudo-iteration'' corresponding to the epilog) and the boxes within a column the instructions executed in the $\pi=2$ time steps until the next iteration starts.
    Above, the instructions are arranged according to the evaluation $f_{1, \texttt{alu1}}((j))$.
    In particular, the instruction with time offset $\tau=2$ \emph{overlaps} with iteration $j+1$, having its condition space shifted from $\{ 0\leq j\leq 5 \}$ to $\{ 1 \leq j \leq 6 \}$ and giving rise to the epilog space $\mathcal{E}=\{ j=6 \}$.
    Below, the instructions are arranged into \keyword{kernels} corresponding to the evaluation $f'_{1, \texttt{alu1}}((j))$ of the folded PST (see main text).
    A \keyword{kernel class} $Q_i$ is a set of instructions issuing the same kernel.%
    %For example, $\mathcal{Q}_2$ is $\mathit{\textcolor{red}{red}}\cap\mathit{\textcolor{green!80!black}{green}}\cap\overline{\mathit{\textcolor{blue}{blue}}}$ (color corresponds to box color).%
  }
  \label{fig:kernel-classes}
\end{figure}

To answer this, we first look at the formation of kernels in a functional unit program, visualized in Figure~\ref{fig:kernel-classes}.
Let $f_{\mathit{pc}, \mathit{fu}}$ be the child of $f_\mathit{pc}$ describing the functional unit program of $\mathit{fu}$.
Then, at the start $t(\vec{J})$ of each iteration $\vec{J}$, the sequence of instructions described by the evaluation $f_{\mathit{pc}, \mathit{fu}}(\vec{J})$ is issued.
Each child $g$ of $f_{\mathit{pc}, \mathit{fu}}$ represents a temporal offset $\tau[g]$ relative to $t(\vec{J})$, from which we compute into which future iterations it overlaps:
Because after $\pi$ timesteps, the next iteration starts, any instruction at offset $\tau$ is executed within slot $(\tau \bmod \pi)$ of the kernel issued in iteration $\operatorname{succ}(\vec{J}, \lfloor\tau / \pi\rfloor)$.
Here, $\operatorname{succ}(\vec{J}, n)$ is the $n$-th successor of $\vec{J}$ according to the intra-tile schedule $\vec{\lambda}^J$.
%In fact, the instruction occupies the slot of the kernel issued in \emph{all} iterations $\vec{J}'\in\mathcal{J}$ that are the $t$-th successor of an iteration $\vec{J}\in\annotate{\mathcal{J}}{g}$.
Since $g$ is issued whenever $\vec{J}$ is in $g$'s condition space $\mathcal{J}_g=\domain(g)$, the instruction at $\annotate{\tau}{g}$ therefore occupies the slot of the kernel issued in \emph{all} iterations that are the $\lfloor \annotate{\tau}{g}/\pi \rfloor$-th successor of an iteration in $\mathcal{J}_g$ (compare the red instructions in Figure~\ref{fig:kernel-classes}), given by
\[
  \operatorname{succ}\left(\mathcal{J}_g, n\right) \coloneqq \{ \vec{J} = \operatorname{succ}(\vec{J}', n) \mid \forall \vec{J}' \in \mathcal{J}_g \}, n = \lfloor\tau[g] / \pi\rfloor.
\]
Using this knowledge, we \keyword{fold} all children of $f_{\mathit{pc}, \mathit{fu}}$ into $\pi$ slots to obtain a tranformed polyhedral syntax tree $f_{\mathit{pc}, \mathit{fu}}'$ that does not describe the sequence of assembly instructions issued at the start of iteration $\vec{J}$, but that instead describes the $\pi$ slots of the \emph{kernel} issued at the start of iteration $\vec{J}$.
The folding operation is elaborated in Algorithm~\ref{alg:folding} and visualized in Figure~\ref{fig:kernel-classes}.

\begin{algorithm}
  \caption{Given $\pi$, fold PST $f_\mathit{fu}$ representing a functional unit program into $f'_\mathit{fu}$}
  \label{alg:folding}
  \begin{algorithmic}
    \State $G\gets \emptyset$
    \For{$g \in \operatorname{children}(f_\mathit{fu})$} \Comment{$g$ represents a time offset $\annotate{\tau}{g}$}
      \State $G' \gets \operatorname{children}(\Call{offset}{g, \lfloor\tau[g] / \pi\rfloor})$
      \State $g' \gets \left( \bigcup_{g''\in G'} \domain(g''), (\tau[g] \bmod \pi), G' \right)$ \Comment {new node with slot index as attribute}
      \State $G \gets G \cup \{ g' \}$
    \EndFor
    \State $f'_\mathit{fu} \gets (\bigcup_{g\in G} \domain(g), \attrs(f), G)$ \Comment{condition space of a parent node is union of children's}

    \Statex
    \Function{offset}{$f, n$} \Comment{recursively offset condition spaces in $f$ by $n$ iterations}
      \If{$\children(f)=\emptyset$} \Comment{if $f$ is a leaf}
        \State \textbf{return} $(\operatorname{succ}(\domain(f), n), \attrs(f), \emptyset)$ \Comment{copy $f$, but with offset condition space}
      \Else
        \State $G \gets \emptyset$
        \For{$g \in \operatorname{children}(f)$} \Comment{offset children condition spaces recursively}
          \State $G \gets G \cup \{ \Call{offset}{g, n} \}$
        \EndFor
        \State \textbf{return} $(\bigcup_{g\in G} \domain(g), \attrs(f), G)$ \Comment{copy $f$, but with offset children}
      \EndIf
    \EndFunction
  \end{algorithmic}
\end{algorithm}

Folding introduces ``iterations'' $\vec{J}$ where $\vec{J}\notin\mathcal{J}$, pseudo-iterations that contain the epilog of the pipelined functional unit program;
we therefore call $\mathcal{E}_g = \mathcal{J}_g \setminus \mathcal{J}$ the \keyword{epilog space} of $g$.
In the following, $\mathcal{J}_{\mathit{pc}, \mathit{fu}}=\mathcal{J}\cup\mathcal{E}_{f'_{\mathit{pc}, \mathit{fu}}}$ denotes intra-tile iteration space including the epilog space and $\mathcal{E}$ the union of all individual epilog spaces.

\begin{example}
After folding $f_{1, \texttt{alu1}}$, we obtain a transformed tree where the second level represents the slot index instead of the time offset $\tau$:
\begin{center}
  \small
  $f'_{1, \texttt{alu1}}:=$
  \begin{forest}
    for tree={grow'=east, child anchor=west, anchor=west, s sep=0pt, tier/.pgfmath=level()}
    [\texttt{alu1}, baseline, anchor=base,
      [\metalabel{slot 0}
        [\texttt{andi}
          [\metalabel{dest}
            [\texttt{od1} \fragmentdomain{1\leq j < p + 1}]
          ]
          [\metalabel{srcA}
            [\texttt{rd0} \fragmentdomain{1 \leq j < p + 1}]
          ]
          [\metalabel{srcB}
            [\texttt{1} \fragmentdomain{1\leq j < p + 1}]
          ]
        ]
      ]
      [\metalabel{slot 1}
        [\texttt{shri}
          [\metalabel{dest}
            [\texttt{od0} \fragmentdomain{j = p - 1}]
            [\texttt{rd1} \fragmentdomain{0\leq j < p - 1}]
          ]
          [\metalabel{srcA}
            [\texttt{rd0} \fragmentdomain{0 \leq j < p}]
          ]
          [\metalabel{srcB}
            [\texttt{1} \fragmentdomain{0\leq j < p}]
          ]
        ]
      ]
    ]
  \end{forest}
\end{center}
For example, the \texttt{andi} instruction, which was originally at \emph{offset} $\tau=2$, now resides at \emph{slot} 0, but with an offset condition space that reflects the overlapping into the next iteration (compare Figure~\ref{fig:kernel-classes}).
This makes the pipelined program's epilog space $\mathcal{E}_{1, \texttt{alu1}} = \{ j=p \}$.
\end{example}

The folded polyhedral syntax tree $f'_{\mathit{pc}, \mathit{fu}}$ of a functional unit gives rise to a set $\mathit{QC}_{\mathit{pc}, \mathit{fu}}$ of \keyword{kernel classes}, that is a partition of $\mathcal{J}\cup\mathcal{E}_{f'_{\mathit{pc}, \mathit{fu}}}$ into subsets $\mathcal{Q}_\mathit{qc}$ of iterations in which the same kernel is issued.
These are determined analogously to processor classes using Algorithm~\ref{alg:processor-classes}.

\begin{example}
For $f'_{1, \texttt{alu1}}$, we obtain 4 kernel classes:
\[
  \mathit{QC}_{1, \texttt{alu1}} = \left\{ \mathcal{Q}_1= \{j = 0\}, \mathcal{Q}_2 = \{ 1\leq j<4 \}, \mathcal{Q}_3 = \{ j=4 \}, \mathcal{Q}_4 = \{ j=5 \} \right\}
\]
The kernel $q_1$ of $\mathcal{Q}_1$, for example, is obtained by evaluating $f'_{1, \texttt{alu1}}$ at any $\vec{J}\in\mathcal{Q}_1$, that is, removing all nodes $g$ where $\vec{J}$ is not in condition space $\domain(g)$:
\begin{center}
  \small
  $f'_{1, \texttt{alu1}}((0))\coloneqq$
  \begin{forest}
    for tree={grow'=east, child anchor=west, anchor=west, s sep=0pt, tier/.pgfmath=level()}
    [\texttt{shift}, baseline, anchor=base,
      [\metalabel{slot 1}
        [\texttt{shri}
          [\metalabel{dest}
            [\texttt{rd1}]
          ]
          [\metalabel{srcA}
            [\texttt{rd0}]
          ]
          [\metalabel{srcB}
            [\texttt{1}]
          ]
        ]
      ]
    ]
  \end{forest}
\end{center}
Converted into an instruction sequence, this evaluation yields (slot index in brackets)
{\small
\begin{verbatim}
  [0] nop
  [1] shri rd1 rd0 1
\end{verbatim}}
(Note that slot 0 has no associated node in $f'_{1, \texttt{alu1}}$---we assume an implied \texttt{nop} in such cases.)
\end{example}

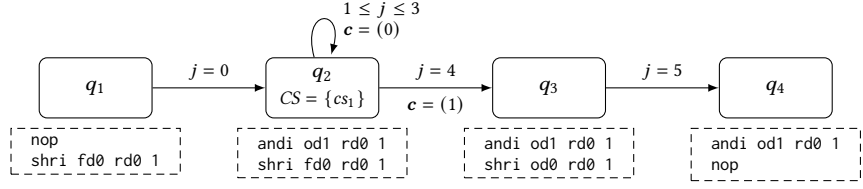
\begin{figure}
  \centering
  \input{figures/tikz/cfg-1}
  \caption{Control flow graph for processor class $\mathcal{P}_1=\{k < 2\}$ and functional unit \texttt{alu1} in the running example.
    Each node represents a kernel, and each edge is annotated with its transition space $\mathcal{T}$, that is, in which iterations the branch represented by the edge is taken.
    For kernels with more than one outgoing edge, the global control signals $\mathit{CS}$ and assignments $\vec{c}$ are also annotated (Section~\ref{sec:control-signal-allocation}).
  }
  \label{fig:cfg}
\end{figure}

Finally, for each processor class and functional unit, the control flow graph $\mathit{CFG}_{\mathit{pc}, \mathit{fu}}$ is constructed from the set of kernel classes $\mathit{QC}_{\mathit{pc}, \mathit{fu}}$ using Algorithm~\ref{alg:cfg}.
The algorithm inserts a node for each kernel class and an edge $e$ between each pair of kernel classes $\mathcal{Q}_i$ and $\mathcal{Q}_j$ where control flow passes from $\mathcal{Q}_i$ to $\mathcal{Q}_j$.
The edge $e$ is annotated with the transition space $\mathcal{T}$, that is, the set of iterations $\vec{J}$ where control flow passes from $\mathcal{Q}_i$ to $\mathcal{Q}_j$.
Figure~\ref{fig:cfg} shows the CFG for processor class $\mathcal{P}_1$ and functional unit \texttt{alu1} of the running example.

\begin{algorithm}
  \caption{Generate control flow graph $\mathit{CFG}$ from kernel classes $\mathit{QC}$ and PST $f'$}
  \label{alg:cfg}
  \begin{algorithmic}
    \State $V \gets \{ v_1, v_2, \ldots, v_{|\mathit{QC}|} \}, E \gets \emptyset$ \Comment{one node for each kernel class}
    \For{$\mathcal{Q}_i \in \mathit{QC}$}
      \State $\annotate{q}{v_i} \gets f'(\mathcal{Q}_i)$, $\annotate{\mathcal{Q}}{v_i} \gets \mathcal{Q}_i$ \Comment{annotate kernel and kernel class to node}
      \For{$\mathcal{Q}_j \in \mathit{QC}$} \Comment{(note: self edges represent repetition)}
        \State $\mathcal{T} \gets \mathcal{Q}_j \cap \operatorname{succ}(\mathcal{Q}_i, 1)$ \Comment{all iterations in $\mathcal{Q}_i$ that have a successor in $\mathcal{Q}_j$}
        \If{$\mathcal{T} \neq \emptyset$} \Comment{if control flow passes from $\mathcal{Q}_i$ to $\mathcal{Q}_j$ in any iteration}
          \State $E\gets E \cup \{ e=(v_i, v_j) \}$, $\annotate{\mathcal{T}}{e} \gets \mathcal{T}$ \Comment{insert edge and annotate transition space}
        \EndIf
      \EndFor
    \EndFor
    \State $\mathit{CFG} \gets (V, E)$
  \end{algorithmic}
\end{algorithm}

The constructed control flow graph represents the functional unit program to be generated:
In any iteration $\vec{J}\in\mathcal{J}_{\mathit{pc}, \mathit{fu}}$, there is exactly one node $v$ where $\vec{J}\in\annotate{\mathcal{Q}}{v}$, representing the kernel $\annotate{q}{v}$ to be issued.
Among its outgoing edges, there is exactly one edge $e=(v, w)$ where $\vec{J}\in\annotate{\mathcal{T}}{e}$, otherwise it is the last node.
Node $w$ represents the kernel $\annotate{q}{w}$ issued in the next iteration.
Hence, the outgoing edges of $v$ represent the set of branch targets and the transition spaces the branch conditions.\footnote{Algorithm~\ref{alg:cfg} generates CFGs with arbitrary outdegree, that is, an arbitrary number of branch targets.
While \acp{TCPA} support a configurable number of simultaneous branch targets, it is usually set to 2 or other low numbers.
Therefore, the outdegree of the control flow graph must be reduced accordingly.
For lack of space, we refer to \cite{boppu2015}.}
The next step is to encode these branch conditions with a set of control signals.

%%%%%%%%%%%%%%%%%%%%%%%%%%%%%%%%%%%%%%%%%%%%%%%%%%%%%%%%%%%%%%%%%%%%%%%%%%%%%%%%%%%%%%%%%%%%%%%%%%%

\subsection{Control signal allocation}
\label{sec:control-signal-allocation}

As stated above, given a $\mathit{CFG}_{\mathit{pc}, \mathit{fu}}=(V, E)$, for each iteration $\vec{J}\in\mathcal{J}_{\mathit{pc}, \mathit{fu}}$, there is exactly one node $v\in V$ where $\vec{J}\in\annotate{\mathcal{Q}}{v}$ with $\operatorname{deg^+}(v)$ branch targets.
Only one target $w$ will be branched to:
the one where $\vec{J}\in\mathcal{T}[(v, w)]$.
Consequently, some entity---in case of a \ac{TCPA} the global controller---must track the current iteration and signal to the \acp{PE} for which edges $\vec{J}$ is in $\mathcal{T}$ in order to select to which target to branch.
For that, the global controller generates (binary) control signals.
\begin{definition}
A \keyword{control signal} is a function
\[
  \mathit{cs}(\vec{J})\colon \mathcal{J}\cup\mathcal{E} \mapsto \{ 0, 1, - \}
\]
that maps an intra-tile iteration $\vec{J}$ to 0, 1, or don't-care (represented by $-$).
We call a control signal \keyword{partial} if it maps at least one $\vec{J}$ to $-$.
\end{definition}

For each node $v\in V$, there are $N_v = \lceil\log_2 \operatorname{deg^+}(v)\rceil$ control signals required to encode the $\operatorname{deg^+}(v)$ outgoing edges of $v$, each of which is given an assignment $\annotate{\vec{c}}{e}\gets (c_1, \ldots, c_{N_v})$ with $c_i\in\{ 0, 1, - \}$ such that these assignments do not overlap for any two outgoing edges.
These assignments can, for example, be determined using binary decision diagrams \cite{boppu2015}.
From these assignments, for each node $v$, we build $N_v$ partial control signals:
\[
  \forall v\in V, 1\leq i \leq N_v\colon
  \mathit{cs}_{v, i}(\vec{J}) := \begin{cases}
    1 & \exists e=(v, w)\in E\colon c_i[e] = 1 \wedge \vec{J}\in \mathcal{T}[e]\\
    0 & \exists e=(v, w)\in E\colon c_i[e] = 0 \wedge \vec{J}\in \mathcal{T}[e]\\
    - & \text{else}
  \end{cases}.
\]

\begin{example}
Only node $q_2$ in Figure~\ref{fig:cfg} requires a control signal because it is the only node with more than one outgoing edge.
We give its two outgoing edges the assignments $\annotate{\vec{c}}{(q_2, q_2)}\gets(0)$ and $\annotate{\vec{c}}{(q_2, q_3)}\gets(1)$.
\end{example}

Across all control flow graphs $\mathit{CFG}_{\mathit{pc}, \mathit{fu}}$, there is now a large set of partial control signals $\mathit{cs}_{\mathit{pc}, \mathit{fu}, v, i}$ \emph{local} to the corresponding node.
Their number usually exceeds the maximum number $C$ of control signals supported by the global controller, which we call \keyword{global} control signals.
We therefore combine the local, partial control signals using an interference graph where each control signal $\mathit{cs}_{\mathit{pc}, \mathit{fu}, v, i}$ is a node and two control signals $\mathit{cs}_1$ and $\mathit{cs}_2$ interfere if
\[
  \exists \vec{J}\in\mathcal{J}\cup\mathcal{E} \colon \mathit{cs}_1(\vec{J}) \neq \mathit{cs}_2(\vec{J})
    \wedge \mathit{cs}_1(\vec{J}) \neq -
    \wedge \mathit{cs}_2(\vec{J}) \neq -.
\]
A $C$-coloring of the vertices then corresponds to the allocation of $C$ global control signals $\mathit{cs}_i$.
For each $\mathit{cs}_i$, the global controller is configured with its \keyword{one domain}, that is, the subset of $\mathcal{J}_{\mathit{pc}, \mathit{fu}}$ where $\mathit{cs}_i(\vec{J})=1$ (for all other iterations 0 is output).
Additionally, each node $v$ in each control flow graph is annotated with the $N_v$-tuple $\annotate{\mathit{CS}}{v}$ of global control signals that were allocated for the node's local control signals;
this information is necessary for program generation.

\begin{example}
For node $q_2$ in Figure~\ref{fig:cfg}, a global control signal $\mathit{cs}_1$ is allocated and configured:
\[
  \mathit{cs}_1(\vec{J}=(j)) := \begin{cases}
    1 & \text{if } j = 4 \\
    0 & \text{else}
  \end{cases}.
\]
That means that if $\mathit{cs}_1(\vec{J})$ is $0$, the next kernel is again $q_2$, if it is $1$, which is only in iteration $j=4$, the next kernel is $q_3$.
\end{example}

Note that for instantiation at runtime, fast heuristics such as greedy graph coloring are sensible because graph coloring is NP-complete.
%Since different functional units tend to branch simultaneously---they are part of the same overarching program, after all---, this results in a reasonably small number of global control signals.
The generated control flow graphs now contain all information necessary for program generation.

%%%%%%%%%%%%%%%%%%%%%%%%%%%%%%%%%%%%%%%%%%%%%%%%%%%%%%%%%%%%%%%%%%%%%%%%%%%%%%%%%%%%%%%%%%%%%%%%%%%

\subsection{Program generation}
\label{sec:program-generation}

Each processor class $\mathcal{P}_\mathit{pc}$ requires the generation of one \ac{PE} program, which is simply a container for the functional unit programs in that processor class.
Generating a \ac{PE} program therefore requires generating the program for each functional unit from its control flow graph $\mathit{CFG}_{\mathit{pc}, \mathit{fu}}$.

In orthogonal instruction processing (see Section~\ref{sec:processing-elements}), each instruction in a functional unit program is a pair of a functional instruction, specifying the operation, and a branch instruction, specifying the next instruction.
While the functional instructions are explicit in the kernels $\annotate{q}{v}$ annotated to the nodes in $\mathit{CFG}_{\mathit{pc}, \mathit{fu}}$, corresponding branch instructions remain to be generated.
The instructions in slots $0\ldots \pi-2$ of a kernel $\annotate{q}{v}$ are each combined with a \texttt{next} branch instruction because each kernel is executed atomically and in order.
However, for the last instruction, the one in slot $\pi-1$, a conditional multi-target branch instruction must be generated that selects the target according to the allocated global control signals $\annotate{\mathit{CS}}{v}=\{ \mathit{cs}_1, \ldots, \mathit{cs}_{N_v}\}$ and the assigned values $\annotate{\vec{c}}{e}$ for all outgoing edges $e\in E^+(v)$.
The subsequent concatenation of all kernels (starting with $\annotate{q}{v}$ where $v$ is the start node, that is, has no incoming edges) yields the functional unit program.
Algorithm~\ref{alg:program-generation} summarizes these two steps.

\begin{algorithm}
  \caption{Generate functional unit program from annotated $\mathit{CFG}=(V, E)$}
  \label{alg:program-generation}
  \begin{algorithmic}
    \State $\mathit{program} \gets []$
    \For{$v \in \operatorname{topological\_sort}(V)$} \Comment{begin with start node}
      \State $\mathit{targets} \gets \{ w \mid (v, w)\in E\}$
      \For{$\mathit{slot} \coloneqq 0$ \textbf{to} $\pi-1$}
        %$\mathit{instruction} \in \annotate{q}{v}$} \Comment{in execution order}
        \If{$\mathit{slot} = \pi - 1$} \Comment{if it is the last instruction in the kernel}
          \State $\mathit{branch} \gets$ \Call{make\_branch}{$v, \mathit{targets}$} \Comment{make an explicit branch to the next kernels}
        \Else
          \State $\mathit{branch} \gets \texttt{next}$ \Comment{otherwise, go unconditionally to the next instruction}
        \EndIf
        \State $\operatorname{append}(\mathit{program}, (\annotate{q}{v}.\mathit{instruction}[\mathit{slot}], \mathit{branch}))$
        %\State \Comment{append tuple of functional and branch instruction}
      \EndFor
    \EndFor

    \Statex
    \Function{make\_branch}{$v, \mathit{targets}$}
      \If{$|\mathit{targets}| = 0$}
        \State \textbf{return} \texttt{halt} \Comment{last node $\rightarrow$ halt execution}
      \ElsIf{$|\mathit{targets}| = 1$}
        \State \textbf{return} \texttt{goto} $\mathit{target}$ \Comment{$\mathit{target}$ is the single element of $\mathit{targets}$}
      \ElsIf{$|\mathit{targets}| > 1$}
        \State \textbf{return} \texttt{if} $\annotate{cs_1}{v}, \ldots, \annotate{cs_{N_v}}{v}$ \texttt{jmp} $\operatorname{target}(\mathit{targets}, 2^{N_v}-1), \ldots, \operatorname{target}(\mathit{targets}, 0)$
        \State\Comment{\textit{$\operatorname{target}(\mathit{targets}, i)$ gives $w$ such that control signal assignment of edge $(v, w)$ matches $i$}}
      \EndIf
    \EndFunction
  \end{algorithmic}
\end{algorithm}

\begin{example}
We obtain the following functional unit program for functional unit \texttt{alu1} in processor class $\mathcal{P}_1$ (the comments show some applicable optional simplifications):
{\small\begin{verbatim}
  q1: nop / next
      shri rd1 rd0 1 / goto q2    // can be simplified to 'next'
  q2: andi od0 rd1 1 / next
      shri rd1 rd0 1 / if ic0 jmp q3, q2
  q3: andi od0 rd1 1 / next
      shri od1 rd0 1 / goto q4    // can be simplified to 'next'
  q4: andi od0 rd1 1 / next
      nop / halt                  // can be merged into previous instruction\end{verbatim}}
The programs for \texttt{alu0} (not shown) and \texttt{alu1} together form the \ac{PE} program for $\mathcal{P}_1$.
\end{example}

%%%%%%%%%%%%%%%%%%%%%%%%%%%%%%%%%%%%%%%%%%%%%%%%%%%%%%%%%%%%%%%%%%%%%%%%%%%%%%%%%%%%%%%%%%%%%%%%%%%

\subsection{I/O access instantiation}
\label{sec:io-access-generation}

The last step is the instantiation of configuration data for the I/O buffers from the access mappings.
Recall that an access mapping $a=(\mathit{reg}, x, \alpha, \mathcal{A})$ maps accesses to $\mathit{reg}$ within iterations $\vec{I}\in\mathcal{A}$ to $x[\alpha(\vec{I})]$.
We call a particular access in an iteration $\vec{I}$ an \keyword{access instance} $a[\vec{I}]$.

\begin{example}
  The access mapping $a_\mathit{bits}$ maps all write accesses to \texttt{od0} with $i<N$ to $\mathit{bits}[Q_\mathit{bits}\vec{I}+\vec{d}_\mathit{bits}]$, that is, $\mathit{bits}[i]$.
  These write accesses correspond to instruction \texttt{andi\,od0\,rd0\,1} in the previous example:
  each time \texttt{od0} is written, the value is stored in $\mathit{bits}[i]$.
\end{example}
  
Now, tiling distributes access instances $a[(\vec{J}, \vec{K})^T]$ in $\mathcal{A}^*\subseteq\mathcal{K}_a\oplus\mathcal{J}$ across multiple \acp{PE} $\mathcal{K}_a$, possibly making them concurrent since the \acp{PE} run in parallel.
Consequently, for each $\vec{K}\in\mathcal{K}_a$ of each access mapping $a\in A$, two parts must be instantiated:
\begin{enumerate}
  \item A connection between a memory bank and the port corresponding to $\mathit{reg}$ of \ac{PE} $\vec{K}$, which entails finding a free bank and a route between the bank and the \ac{PE} on the interconnect.
  %However, systolic algorithms usually access external data only in border \acp{PE}, making it possible to forego routing and connect each border \ac{PE} directly to a memory bank.
  \item The configuration of the allocated memory bank's address generator, consisting of the coefficients of an affine address function derived from $\alpha$ and the memory layout of $x$ that maps the intra-tile iteration $\vec{J}$ to an address, and the intra-tile domain $\widehat{\mathcal{J}}=\operatorname{split}(\vec{K}, \mathcal{A}^*)$ in the access space, required to generate an enable signal.
\end{enumerate}

\begin{figure}
  \centering
  \input{figures/tikz/channel-route}
  \caption{Concrete interconnect wrapper and memory bank configuration for the running example.
    The propagation channels are replicated across the $t=3$ \acp{PE} and each is connected to memory banks according to the access mappings $a_\mathit{in}$ (west bank) and $a_\mathit{bits}$ (north banks).
    Each bank is annotated with the iterations when the enable signal is 1 (first line) and which element of the associated array is accessed (second line).
  }
  \label{fig:config-overview}  
\end{figure}
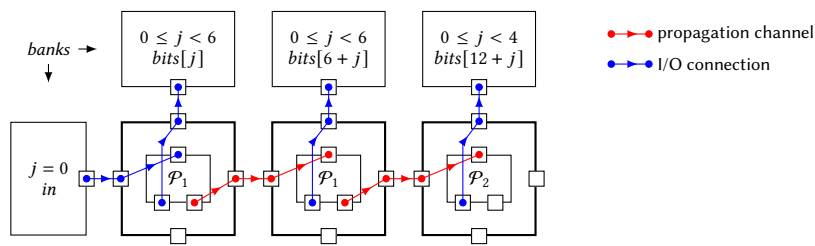

This clearly makes the time complexity of this step linear in the number of involved \acp{PE} $|\mathcal{K}_a|$ in the general case.
However, if I/O variables are only accessed at the borders, routing becomes unnecessary---the border \acp{PE} have a direct connection to the I/O banks.
This allows the compiler to generate the above two parts already at compile time.
Figure~\ref{fig:config-overview} shows the interconnect and I/O buffers parts of the instantiated configuration for the running example.
%(Note that buffer banks usually cannot fit the entire required data, which means they need to be filled or drained periodically to avoid stalling, which is managed by a I/O controller.)

%\subsection{Summary}

%After instantiation, the given symbolic configuration and parameter value assignments have been used to generate a concrete configuration that can be loaded onto the target \ac{TCPA} and executed.
%Instantiation only contains polynomial-time algorithms (with the exception of control signal allocation, which can be solved greedily), program instantiation is independent of the number of allocated \acp{PE}, and I/O access instantiation is independent of the number of allocated \acp{PE} if it can already be solved at compile time, making runtime instantiation viable.
%Keeping time complexity low serves to reduce the time it takes at runtime to amortize the time required for instantiation.

%% file: figures/tikz/processor-classes.tex
\colorlet{iteration1}{blue}
\colorlet{iteration2}{green}
\colorlet{iteration3}{red}
\colorlet{iteration4}{magenta}
\colorlet{iteration5}{cyan}

\begin{tikzpicture}[
  iteration/.style={circle,draw,fill=white,inner sep=0pt,minimum size=0.25cm, font=\ttfamily\tiny},
  tight background,
]

  \foreach \k in {0, 1, 2} {
    \foreach \j in {0, 1, ..., 5} {
      \pgfmathsetmacro\i{int(6*\k+\j)}
      \pgfmathsetmacro\x{2.625*\k+0.375*\j}
      \ifthenelse{\k<2 \AND \j=0}{
        \node[iteration, fill=iteration1!30] (i-\k-\j) at (\x, 0) {};
      }{}
      \ifthenelse{\k=2 \AND \j=0}{
        \node[iteration, fill=iteration4!30] (i-\k-\j) at (\x, 0) {};
      }{}
      \ifthenelse{\k<2 \AND \j=5}{
        \node[iteration, fill=iteration3!30] (i-\k-\j) at (\x, 0) {};
      }{}
      \ifthenelse{\k<2 \AND \j>0 \AND \j<5}{
        \node[iteration, fill=iteration2!30] (i-\k-\j) at (\x, 0) {};
      }{}
      \ifthenelse{\k=2 \AND \j>0 \AND \j<4}{
        \node[iteration, fill=iteration5!30] (i-\k-\j) at (\x, 0) {};
      }{}
      \ifthenelse{\k=2 \AND \j>3}{
        \node[iteration, densely dotted] (i-\k-\j) at (\x, 0) {};
      }{}

      \ifthenelse{\k<2}{
        \node[iteration] (domain-1-\i) at (\x, -0.75) {1};
      }{
        \node[iteration, draw=none, fill=gray, text=white] (domain-1-\i) at (\x, -0.75) {0};
      }
      \ifthenelse{\k=2 \AND \j<4}{
        \node[iteration] (domain-2-\i) at (\x, -1.25) {1};
      }{
        \ifthenelse{\k=2}{
          \node[iteration, draw=none, fill=gray, text=white] (domain-2-\i) at (\x, -1.25) {0};
        }{
          \node[iteration, draw=none, fill=gray, text=white] (domain-2-\i) at (\x, -1.25) {0};
        }
      }

      \ifthenelse{\j=0}{
        \node[iteration] (domain-3-\i) at (\x, -1.75) {1};
      }{
        \node[iteration, draw=none, fill=gray, text=white] (domain-3-\i) at (\x, -1.75) {0};
      }

      \ifthenelse{\j=5 \AND \k<2}{
        \node[iteration] (domain-4-\i) at (\x, -2.25) {1};
      }{
        \node[iteration, draw=none, fill=gray, text=white] (domain-4-\i) at (\x, -2.25) {0};
      }
    }

    \node[fit={(i-\k-0)(i-\k-5)}, draw] (tile-\k) {};

    \draw[thick, ->, >=latex] ([yshift=0.5cm] i-\k-0.center)
      -- ([shift={(0.25, 0.5)}] i-\k-5.center) node[right, font=\footnotesize] {$j$};
    \foreach \j in {0, 1, ..., 5} {
      \draw ([yshift=0.375cm] i-\k-\j.center) -- ++(up:0.25cm) node[above, font=\footnotesize] {$\j$};
    }    
  }

  \draw[thick, ->, >=latex] ([yshift=1.125cm] i-0-0.center)
    -- ([shift={(0.25, 1.125)}] i-2-5.center) node[right, font=\footnotesize] {$k$};
  \foreach \k in {0, 1, 2} {
    \draw ([yshift=1cm] i-\k-0.center)
      -- ++(up:0.25cm) node[above, font=\footnotesize] {$\k$};
  }

  \node[left, inner sep=2pt, font=\footnotesize] at (-0.5, -0.75) (domain-1-label) {$\widehat{\mathcal{I}}_1=(\{ j < 6 \}, \{ k<2 \})$};
  \node[left, inner sep=2pt, font=\footnotesize] at (-0.5, -1.25) (domain-2-label) {$\widehat{\mathcal{I}}_2=(\{ j < 4 \}, \{ k=2 \})$};
  \node[left, inner sep=2pt, font=\footnotesize] at (-0.5, -1.75) (domain-3-label) {$g_2\colon \widehat{\mathcal{I}}_3=(\{ j = 0 \}, \{ k \leq 2 \})$};
  \node[left, inner sep=2pt, font=\footnotesize] at (-0.5, -2.25) (domain-4-label) {$g_4\colon \widehat{\mathcal{I}}_4=(\{ j = 5 \}, \{ k < 2 \})$};

  \draw[decorate,decoration={brace,amplitude=3pt,mirror}]
    (domain-1-label.north west) -- (domain-1-label.north west |- domain-2-label.south west)
      node[midway, left, xshift=-0.25em, font=\footnotesize] {$g_1\colon$};

  \begin{scope}[on background layer]
    \node[fill=gray!30, inner sep=2pt, fit={(domain-1-0) (domain-1-5)}] (p-1-0) {};
    \node[fill=gray!30, inner sep=2pt, fit={(domain-1-6) (domain-1-11)}] (p-1-1) {};

    \node[fill=gray!30, inner sep=2pt, fit={(domain-2-12) (domain-2-17)}] (p-2-2) {};

    \node[fill=gray!30, inner sep=2pt, fit={(domain-3-0) (domain-3-5)}] (p-3-0) {};
    \node[fill=gray!30, inner sep=2pt, fit={(domain-3-6) (domain-3-11)}] (p-3-1) {};
    \node[fill=gray!30, inner sep=2pt, fit={(domain-3-12) (domain-3-17)}] (p-3-2) {};

    \node[fill=gray!30, inner sep=2pt, fit={(domain-4-0) (domain-4-5)}] (p-4-0) {};
    \node[fill=gray!30, inner sep=2pt, fit={(domain-4-6) (domain-4-11)}] (p-4-1) {};
  \end{scope}

  \node[draw, gray!30, very thick, inner sep=0.1825cm, fit={(domain-1-0) (domain-4-5)}] (pc-0) {};
  \node[draw, gray!30, very thick, inner sep=0.1825cm, fit={(domain-1-6) (domain-4-11)}] (pc-1) {};
  \node[draw, gray!30, very thick, inner sep=0.1825cm, fit={(domain-1-12) (domain-4-17)}] (pc-2) {};
  \node[inner sep=0pt, below=2pt of pc-0.south] {$\mathcal{P}_1$};
  \node[inner sep=0pt, below=2pt of pc-1.south] {$\mathcal{P}_1$};
  \node[inner sep=0pt, below=2pt of pc-2.south] {$\mathcal{P}_2$};

  \foreach \i in {0, 6} {
    \pgfmathsetmacro\k{int(\i/6)}
    \pgfmathsetmacro\j{int(mod(\i, 6))}
    \node[draw, iteration1, rounded corners=4pt, inner sep=2pt,
      fit={(domain-1-\i.north west) (domain-4-\i.south east)}] {};
  }

  \foreach \i in {12} {
    \pgfmathsetmacro\k{int(\i/6)}
    \pgfmathsetmacro\j{int(mod(\i, 6))}
    \node[draw, iteration4!80!black, rounded corners=4pt, inner sep=2pt,
      fit={(domain-1-\i.north west) (domain-4-\i.south east)}] {};
  }

  \foreach \i in {1, 2, 3, 4, 7, 8, 9, 10} {
    \pgfmathsetmacro\k{int(\i/6)}
    \pgfmathsetmacro\j{int(mod(\i, 6))}
    \node[draw, iteration2!70!black, rounded corners=4pt, inner sep=2pt,
      fit={(domain-1-\i.north west) (domain-4-\i.south east)}] {};
  }

  \foreach \i in {13, 14, 15} {
    \pgfmathsetmacro\k{int(\i/6)}
    \pgfmathsetmacro\j{int(mod(\i, 6))}
    \node[draw, iteration5!80!black, rounded corners=4pt, inner sep=2pt,
      fit={(domain-1-\i.north west) (domain-4-\i.south east)}] {};
  }

  \foreach \i in {5, 11} {
    \pgfmathsetmacro\k{int(\i/6)}
    \pgfmathsetmacro\j{int(mod(\i, 6))}
    \node[draw, iteration3, rounded corners=4pt, inner sep=2pt,
      fit={(domain-1-\i.north west) (domain-4-\i.south east)}] {};
  }

  \foreach \i in {16, 17} {
    \pgfmathsetmacro\k{int(\i/6)}
    \pgfmathsetmacro\j{int(mod(\i, 6))}
    \node[draw, black, rounded corners=4pt, densely dotted, inner sep=2pt,
      fit={(domain-1-\i.north west) (domain-4-\i.south east)}] {};
  }

\end{tikzpicture}

%% file: figures/tikz/kernel-classes.tex
\begin{tikzpicture}[
  font=\small\sffamily,
  iteration/.style={circle,draw,fill=white,inner sep=0pt,minimum size=0.25cm},
  slot/.style={draw, inner sep=0pt, minimum height=0.325cm, minimum width=1.6cm, font=\ttfamily\scriptsize},
  tight background,
]

  \foreach \tau in {0, ..., 2} {
      \foreach \j in {0, ..., 6} {
        \node[slot, draw=none, fill=none] (tau-\tau-\j) at ($1.7*(\j, 0)+0.425*(0, \tau)+(0,1.75)$) {};
      }
  }

  \foreach \slot in {0, ..., 1} {
      \foreach \j in {0, ..., 6} {
        \node[slot, draw=none, fill=none] (slot-\slot-\j) at ($1.7*(\j, 0)+0.425*(0, \slot)$) {};
      }
  }

  %\node[fill=black, inner sep=1pt, fit={(tau-0-0.south west) (tau-1-0.north east)}] {};
  %\node[fill=black, inner sep=1pt, fit={(tau-2-1.south west) (tau-2-1.north east)}] {};

  %\node[fill=gray!30, inner sep=1pt, fit={(tau-0-2.south west) (tau-1-2.north east)}] {};
  %\node[fill=gray!30, inner sep=1pt, fit={(tau-2-3.south west) (tau-2-3.north east)}] {};

  \draw[<->, >=latex]
    (tau-2-0.west) -- node[midway, fill=white, font=\scriptsize] {$\lfloor \tau/\pi \rfloor$} (tau-2-1.west);

  \foreach \j in {0, ..., 5} {
    \node[slot, densely dotted, text=lightgray, font=\sffamily\itshape\scriptsize] (tau-0-\j-slot) at (tau-0-\j) {};
  }

  \node[slot, fill=green!30] (tau-1-0-slot) at (tau-1-0) {shri\,fd0\,rd0\,1};
  \node[slot, fill=green!30] (tau-1-1-slot) at (tau-1-1) {shri\,fd0\,rd0\,1};
  \node[slot, fill=green!30] (tau-1-2-slot) at (tau-1-2) {shri\,fd0\,rd0\,1};
  \node[slot, fill=green!30] at (tau-1-3) {shri\,fd0\,rd0\,1};
  \node[slot, fill=green!30] at (tau-1-4) {shri\,fd0\,rd0\,1};
  \node[slot, fill=blue!30] at (tau-1-5) {shri\,od1\,rd0\,1};

  \node[slot, draw=none, fill=none] (tau-2-0-slot) at (tau-2-0) {};
  \node[slot, fill=red!30] (tau-2-1-slot) at (tau-2-1) {andi\,od0\,rd0\,1};
  \node[slot, fill=red!30] at (tau-2-2) {andi\,od0\,rd0\,1};
  \node[slot, fill=red!30] (tau-2-3-slot) at (tau-2-3) {andi\,od0\,rd0\,1};
  \node[slot, fill=red!30] at (tau-2-4) {andi\,od0\,rd0\,1};
  \node[slot, fill=red!30] at (tau-2-5) {andi\,od0\,rd0\,1};
  \node[slot, fill=red!30] at (tau-2-6) {andi\,od0\,rd0\,1};

  \foreach \j in {0, ..., 6} {
    \pgfmathsetmacro\t{int(2*\j)}
    \pgfmathsetmacro\tn{int(2*\j+1)}
    \node[above=0pt of tau-2-\j] {$j=\j$};
  }

  \foreach[evaluate=\j as \nj using {int(\j+1)}] \j in {0, ..., 5} {
    %\draw[thick]
      %([xshift=-0.075cm] tau-0-\j.north east) -- ([xshift=-0.075cm] tau-1-\j.south east)
      %(tau-1-\j.north east) -- (tau-2-\nj.south west);
    %\draw[thick]
      %(tau-0-\j.east) circle (0.025cm)
      %-- (tau-1-\j.east) circle (0.025cm)
      %-- (tau-1-\j.north east)
      %-- (tau-2-\nj.south west)
      %-- (tau-2-\nj.west) circle (0.025cm);
    \draw[thick]
      (tau-0-\j.south east)
      -- (tau-1-\j.north east)
      -- (tau-2-\nj.south west)
      -- (tau-2-\nj.south east)
      -- (tau-2-\nj.north east);
    \node[below=2pt of tau-0-\j, inner sep=0pt, font=\footnotesize] {$f_{1, \texttt{alu1}}((\j))$};
  }

  %\draw[densely dashed, gray]
    %($(tau-0-0.south east)!.5!(tau-0-1.south west)$) -- ($(tau-1-0.north east)!.5!(tau-2-1.south west)$)
    %-- ($(tau-1-1.north east)!.5!(tau-2-2.south west)$) -- ($(tau-2-1.north east)!.5!(tau-2-2.north west)$);
  
  %\draw[gray]
    %([shift={(2pt, -2pt)}] tau-0-0.south east) -- ([shift={(-2pt, -2pt)}] tau-2-1.south west);
  
  %\draw[] (tau-0-0-slot.south west) rectangle (tau-1-0-slot.north east);
  %\draw[] (tau-2-1-slot.south west) rectangle (tau-2-1-slot.north east);
  %\draw[densely dotted] (tau-1-1-slot.south west) rectangle (tau-1-1-slot.north east);
  %\draw[] (tau-1-2-slot.south west) rectangle (tau-1-2-slot.north east);
  %\draw[] (tau-2-3-slot.south west) rectangle (tau-2-3-slot.north east);

  \begin{scope}[every node/.style={draw}]
    \node[slot] (slot-0-0-slot) at (slot-0-0) {nop};
    \node[slot, fill=red!30] at (slot-0-1) {andi\,od0\,rd0\,1};
    \node[slot, fill=red!30] at (slot-0-2) {andi\,od0\,rd0\,1};
    \node[slot, fill=red!30] at (slot-0-3) {andi\,od0\,rd0\,1};
    \node[slot, fill=red!30] at (slot-0-4) {andi\,od0\,rd0\,1};
    \node[slot, fill=red!30] at (slot-0-5) {andi\,od0\,rd0\,1};
    \node[slot, fill=red!30] at (slot-0-6) {andi\,od0\,rd0\,1};
    \node[slot, fill=green!30] (slot-1-0-slot) at (slot-1-0) {shri\,fd0\,rd0\,1};
    \node[slot, fill=green!30] at (slot-1-1) {shri\,fd0\,rd0\,1};
    \node[slot, fill=green!30] at (slot-1-2) {shri\,fd0\,rd0\,1};
    \node[slot, fill=green!30] at (slot-1-3) {shri\,fd0\,rd0\,1};
    \node[slot, fill=green!30] at (slot-1-4) {shri\,fd0\,rd0\,1};
    \node[slot, fill=blue!30] at (slot-1-5) {shri\,od1\,rd0\,1};
    \node[slot] at (slot-1-6) {nop};    
  \end{scope}

  \foreach \j in {0, ..., 6} {
    \node[below=2pt of slot-0-\j, inner sep=0pt, font=\footnotesize] (slot-\j-label) {$f'_{1, \texttt{alu1}}((\j))$};
  }

  \draw[decoration={brace, mirror},decorate]
    (slot-1-label.south west) -- node[midway, below, yshift=-1pt, font=\footnotesize] {$\mathcal{Q}_2$} (slot-4-label.south east);
  \draw[decoration={brace, mirror},decorate]
    (slot-0-label.south west) -- node[midway, below, yshift=-1pt, font=\footnotesize] {$\mathcal{Q}_1$} (slot-0-label.south east);
  \draw[decoration={brace, mirror},decorate]
    (slot-5-label.south west) -- node[midway, below, yshift=-1pt, font=\footnotesize] {$\mathcal{Q}_3$} (slot-5-label.south east);
  \draw[decoration={brace, mirror},decorate]
    (slot-6-label.south west) -- node[midway, below, yshift=-1pt, font=\footnotesize] {$\mathcal{Q}_4$} (slot-6-label.south east);

  \draw[decoration={brace, raise=2pt},decorate]
    (slot-1-6.north west) -- node[midway, above, yshift=3pt, font=\footnotesize] {$\mathcal{E}$} (slot-1-6.north east);
  \draw[decoration={brace, raise=2pt},decorate]
    (slot-1-0.north west) -- node[midway, above, yshift=3pt, font=\footnotesize] {$\mathcal{J}$} (slot-1-5.north east);

  %\node[below=0pt of slot-0-1, font=\footnotesize] {$\mathcal{Q}_2$};
  %\node[below=0pt of slot-0-2, font=\footnotesize] {$\mathcal{Q}_2$};
  %\node[below=0pt of slot-0-3, font=\footnotesize] {$\mathcal{Q}_2$};
  %\node[below=0pt of slot-0-4, font=\footnotesize] {$\mathcal{Q}_2$};
  %\node[below=0pt of slot-0-5, font=\footnotesize] {$\mathcal{Q}_3$};
  %\node[below=0pt of slot-0-6, font=\footnotesize] {$\mathcal{Q}_4$};

  \node[left] (tau-0-label) at (tau-0-0-slot.west) {$\tau=0$};
  \node[left] (tau-1-label) at (tau-1-0-slot.west) {$\tau=1$};
  \node[left] (tau-2-label) at (tau-2-0-slot.west) {$\tau=2$};

  \node[left] (slot-0-label) at (slot-0-0-slot.west) {\textit{slot} $0$};
  \node[left] (slot-1-label) at (slot-1-0-slot.west) {\textit{slot} $1$};

  \draw[->, >=latex]
    (tau-2-label.west) -- ++(left:0.75) coordinate (temp) -- ([yshift=-2pt] temp |- slot-0-label.west) -- ([yshift=-2pt] slot-0-label.west);
  \draw[->, >=latex]
    (tau-0-label.west) -- ++(left:0.25) coordinate (temp) -- ([yshift=1pt] temp |- slot-0-label.west) -- ([yshift=1pt] slot-0-label.west);
  \draw[->, >=latex]
    (tau-1-label.west) -- ++(left:0.5) coordinate (temp) -- node[midway, yshift=-4pt, xshift=2pt, fill=white, font=\footnotesize] {$\tau\bmod\pi$} (temp |- slot-1-label.west) -- (slot-1-label.west);

\end{tikzpicture}

%% file: figures/tikz/cfg-1.tex
\begin{tikzpicture}[
  cfg node/.style={
    draw, rectangle, rounded corners=3pt,
    inner sep=0pt,
    minimum width=1.5cm, minimum height=0.8cm,
    font=\small,
    align=center,
  },
  cfg edge/.style={->,>=latex},
  cfg edge label/.style={font=\footnotesize},
  kernel/.style={draw, minimum width=2.25cm, densely dashed, font=\footnotesize, align=left},
  node distance=0.125cm and 1.5cm,
]
  \node[cfg node] (k1) {$q_1$};
  \node[cfg node, right=of k1] (k2) {$q_2$\\\footnotesize$\mathit{CS}=\{ \mathit{cs}_1 \}$};
  \node[cfg node, right=of k2] (k3) {$q_3$};
  \node[cfg node, right=of k3] (k4) {$q_4$};

  \path[cfg edge]
    (k1) edge node[cfg edge label, above] {$j = 0$} (k2)
    (k2) edge[loop above] node[cfg edge label, right, xshift=0.5em, align=left] {$1\leq j \leq 3$\\$\vec{c}=(0)$} (k2)
         edge node[cfg edge label, above] {$j=4$} node[cfg edge label, below] {$\vec{c}=(1)$} (k3)
    (k3) edge node[cfg edge label, above] {$j=5$} (k4)
    ;

  \node[below=of k1, kernel] {\texttt{nop}\\\texttt{shri fd0 rd0 1}};
  \node[below=of k2, kernel] {\texttt{andi od1 rd0 1}\\\texttt{shri fd0 rd0 1}};
  \node[below=of k3, kernel] {\texttt{andi od1 rd0 1}\\\texttt{shri od0 rd0 1}};
  \node[below=of k4, kernel] {\texttt{andi od1 rd0 1}\\\texttt{nop}};
\end{tikzpicture}

%% file: figures/tikz/channel-route.tex
\begin{tikzpicture}[
  font=\sffamily\footnotesize,
  wrapper/.style={
    draw, thick,
    minimum size=1.5cm,
  },
  pe/.style={
    font=\sffamily\footnotesize,
    draw,
    minimum width=0.85cm,
    minimum height=0.625cm,
    inner sep=0pt,
  },
  port/.style={
    draw,
    fill=white,
    minimum size=0.2cm,
    inner sep=0pt,
  },
  ns bank/.style={
    draw,
    minimum width=1.5cm, minimum height=1cm,
  },
  we bank/.style={
    draw,
    minimum width=1cm, minimum height=1.5cm,
  },
]

  \foreach \r in {0} {
    \foreach \c in {0, 1, 2} {
      \node[wrapper] (iw-\c-\r) at ($2*(\c, -\r)$) {};
      \foreach \dir in {north, east, south, west} {
        \newcommand\cornera{}
        \newcommand\cornerb{}
        \ifthenelse{\equal{\dir}{north} \OR \equal{\dir}{south}}{
          \renewcommand\cornera{\dir\space west}
          \renewcommand\cornerb{\dir\space east}
        }{% east/west
          \renewcommand\cornera{north \dir}
          \renewcommand\cornerb{south \dir}
        }
        \foreach \i in {0} {
          \pgfmathsetmacro\pos{1/2}
          \node[port] (iw-\c-\r\space port-\dir-\i) at ($(iw-\c-\r.\cornera)!\pos!(iw-\c-\r.\cornerb)$) {};
        }
      }

      \node[pe] (pe-\c-\r) at (iw-\c-\r.center) {};
      \foreach \i in {0} {
        \pgfmathsetmacro\pos{1/2}
        \node[port] (pe-\c-\r\space input-port-\i) at ($(pe-\c-\r.north west)!\pos!(pe-\c-\r.north east)$) {};
      }
      \foreach \i in {0, 1} {
        \pgfmathsetmacro\pos{1/4+\i/2}
        \node[port] (pe-\c-\r\space output-port-\i) at ($(pe-\c-\r.south west)!\pos!(pe-\c-\r.south east)$) {};
      }
    }
  }

  \foreach \i in {0, 1, 2} {
    \node[ns bank] (bank-n-\i) at ($2*(\i, 0.8625)$) {};
    \node[port] (bank-n-\i\space port) at ($(bank-n-\i.south west)!.5!(bank-n-\i.south east)$) {};
  }

  \foreach \i in {0} {
    \node[we bank] (bank-w-\i) at ($2*(-0.8625, \i)$) {};
    \node[port] (bank-w-\i\space port) at ($(bank-w-\i.north east)!.5!(bank-w-\i.south east)$) {};
  }

  \begin{scope}[
    connection/.style={
      -{Circle[length=1mm]}, shorten >=-.5mm,
      decoration={
        markings,
        mark=at position 0.5 with {\arrow[xshift=3pt]{latex}}
      },
      postaction={decorate},  
    },
    both heads/.style={
      {Circle[length=1mm]}-{Circle[length=1mm]}, shorten >=-.5mm, shorten <=-.5mm,
    },  
  ]
    \begin{scope}[red]
      \draw[connection, both heads]
        (pe-0-0 output-port-1.center) -- (iw-0-0 port-east-0.center);
      \draw[connection]
        (iw-0-0 port-east-0.center) -- (iw-1-0 port-west-0.center);
      \draw[connection]
        (iw-1-0 port-west-0.center) -- (pe-1-0 input-port-0.center);

      \draw[connection, both heads]
        (pe-1-0 output-port-1.center) -- (iw-1-0 port-east-0.center);
      \draw[connection]
        (iw-1-0 port-east-0.center) -- (iw-2-0 port-west-0.center);
      \draw[connection]
        (iw-2-0 port-west-0.center) -- (pe-2-0 input-port-0.center);
    \end{scope}
  
    \begin{scope}[
      blue,
      connection 7/.style={
        blue,
        -{Circle[length=1mm]}, shorten >=-.5mm,
        decoration={
          markings,
          mark=at position 0.7 with {\arrow[xshift=3pt]{latex}}
        },
        postaction={decorate},  
      },
    ]
      \draw[connection, both heads]
        (bank-w-0 port.center) -- (iw-0-0 port-west-0.center);
      \draw[connection]
        (iw-0-0 port-west-0.center) -- (pe-0-0 input-port-0.center);
  
      \draw[connection 7, both heads]
        (pe-0-0 output-port-0.center) -- ++(0, 0.8) -- (iw-0-0 port-north-0.center);
      \draw[connection]
        (iw-0-0 port-north-0.center) -- (bank-n-0 port.center);
  
      \draw[connection 7, both heads]
        (pe-1-0 output-port-0.center) -- ++(0, 0.8) -- (iw-1-0 port-north-0.center);
      \draw[connection]
        (iw-1-0 port-north-0.center) -- (bank-n-1 port.center);
  
      \draw[connection 7, both heads]
        (pe-2-0 output-port-0.center) -- ++(0, 0.8) -- (iw-2-0 port-north-0.center);
      \draw[connection]
        (iw-2-0 port-north-0.center) -- (bank-n-2 port.center);
    \end{scope}

    \draw[red, connection, both heads]
      ([shift={(1, 0.2)}] bank-n-2.east) -- ++(right:0.5) node[black, right] {propagation channel};
    \draw[blue, connection, both heads]
      ([shift={(1, -0.2)}] bank-n-2.east) -- ++(right:0.5) node[black, right] {I/O connection};
  \end{scope}

  \node at (pe-0-0) {$\mathcal{P}_1$};
  \node at (pe-1-0) {$\mathcal{P}_1$};
  \node at (pe-2-0) {$\mathcal{P}_2$};

  %\node[below right=1pt and 1pt of bank-n-0.north west] {\textit{Bank}};
  \node[align=center] at (bank-w-0) {$j = 0$\\$\mathit{in}$};
  \node[align=center] at (bank-n-0) {$0\leq j < 6$\\$\mathit{bits}[j]$};
  \node[align=center] at (bank-n-1) {$0\leq j < 6$\\$\mathit{bits}[6+j]$};
  \node[align=center] at (bank-n-2) {$0\leq j < 4$\\$\mathit{bits}[12+j]$};

  \node (banks label) at (bank-w-0 |- bank-n-0) {\textit{banks}};
  \draw[->, >=latex] (banks label.east) -- ++(right:0.25cm);
  \draw[->, >=latex] (banks label.south) -- ++(down:0.25cm);
\end{tikzpicture}

%% file: sections/discussion.tex
\section{Experiments and discussion}
\label{sec:experiments}

In the following, we experimentally show the validity of the two claims given in the beginning:
that time complexity of program instantiation does not directly depend on the number of \acp{PE} and that a symbolic configuration is a space-efficient representation.
%(2) time complexity of I/O access instantiation is at worst linear in the number of \acp{PE} if non-border \acp{PE} access memory, and

In particular, including the running example (bit extraction), we compiled a symbolic configuration according to Section~\ref{sec:symbolic-compilation} for each of the following real-world loop programs:
a pipelined implementation of an FIR filter (2-dimensional loop),
matrix-matrix multiplication (3-dimensional loop),
and a convolutional layer within a CNN (6-dimensional loop).
The choice is based on the intention to cover a variety of both application domains and dimensionality.
For each symbolic configuration, we instantiated, as described in Section~\ref{sec:instantiation}, six concrete configurations corresponding to six tilings:
three resulting in a 1-dimensional region of \acp{PE} (1, 16, and 32 \acp{PE}), and three resulting in a 2-dimensional region ($4\times 4$, $8\times 8$, and $32\times 32$)\footnote{Both loop bounds and tile sizes were chosen appropriately to result in these \ac{PE} regions.}.
For each instantiation run, we measured the execution time of program instantiation and the size of the concrete configuration.

Since we want to show time complexity, we are interested in normalized execution times, summarized in Table~\ref{table:benchmarks}.
Each row represents one of the examples and contains, for each of the six tilings, the execution time of program instantiation normalized to the runtime of program instantiation in the case of a single tile (1 \ac{PE}).
The number of resulting processor classes is given in parentheses.
Clearly, the execution time of program instantiation is roughly linear in the number of processor classes and \emph{not} in the number of \acp{PE}, as is, for example, evident for the matrix multiplication example:
Program instantiation for both $4\times 4=16$ and $32\times 32=1024$ \acp{PE} takes about equally as long because both have two processor classes, meaning two programs need to be instantiated.
The instantiation phase therefore effortlessly scales to the ever-increasing number of \acp{PE}.
(Note that some tilings may result in more complex control flow analysis, as is for example seen in the convolution example, where instantiation for $4\times 4$ \acp{PE} takes about four times as long as for $1\times 16$ \acp{PE}, despite having only double as many processor classes.
However, it is still independent of the number of \acp{PE}.)

%Figure~\ref{table:io-benchmarks} plots the execution times of I/O access instantiation for the matrix multiplication example, which was also normalized to the case of 1 \ac{PE}.
%Since we compiled the symbolic configuration for an implementation without localized outputs, each allocated \acp{PE} performs output accesses, making the time complexity approximately linear in the number of \acp{PE}.
%It is therefore advisable to localize in- and outputs to allow the compiler to derive I/O routes and addresss generator configurations already at compile time.

Table~\ref{table:size-benchmarks} shows the size of each concrete configuration normalized to the size of the symbolic configuration it was instantiated from.
Excluding the bit extraction example, all concrete configurations by themselves were already larger than the symbolic configuration.
Consequently, runtime instantiation significantly saves memory even if only a small number of concrete configurations were necessary at runtime.
For example, storing the symbolic configuration for the CNN example saves about 95\,\% space compared to storing both concrete configurations for $4\times 4$ and $8\times 8$ \acp{PE}.

%\begin{figure}
  %\begin{center}
    %\input{sections/table-buffer-instantiation}
  %\end{center}
  %\caption{Plot of the runtimes of I/O access instantiation of the matrix multiplication example, which performs output accesses in each \ac{PE}, normalized to the case of only allocating 1 \ac{PE}.
    %Clearly, the time complexity is roughly linear in the number of \acp{PE}.
  %}
  %\label{table:io-benchmarks}
%\end{figure}

\subsection{Practical insights}

As proof of concept, we implemented the instantiation phase as described in Section~\ref{sec:instantiation} using \texttt{isl}, the integer set library \cite{isl}, to represent parametric iteration and condition spaces.
This implementation---which was also used for the experiments in the previous section---, is functionally complete, but was not implemented with optimizing performance in mind.
Instead, it uses a high level of abstraction (which includes \texttt{isl}) to aid in verifying correctness of the approach and in analyzing intermediate artifacts.
For employment in an embedded system for instantiation at runtime, a more optimized implementation is desirable;
in particular, profiling showed that a significant part of the execution time was spent during \texttt{isl} calls.
A dedicated, simplified, non-parametric representation of iteration and condition spaces may thus lead to a considerate speed-up.
Furthermore, if the host platform supports multi-threading, the execution time of program instantiation can be significantly improved by instantiating each processor class in parallel.

On a related note, we noticed that control flow graph generation is often the most computationally expensive part of program instantiation, caused by its quadratic time complexity in the number of program blocks.
In a more practical implementation, this complexity should be improved by decreasing the number of successor candidates considered for each program block.
This might be achieved by bringing the program blocks in a clever order according to the loop schedule.

\begin{table}
    \begin{center}
      \input{sections/table_program_generation}
    \end{center}
    \caption{Relative runtimes of program instantiation for a set of mappings of various loop programs.
      Each row represents a symbolic configuration of the listed algorithm and each column the instantiation for one of six tilings, three resulting in a one-dimensional and three in a two-dimensional \ac{PE} allocation.
      For each instantiation, the runtime relative to the runtime of the first column is listed;
      the number of processor classes is listed in parentheses.
      The table clearly shows that the time complexity of instantiation is roughly proportional to the number of processor classes and \emph{not} to the number of processing elements.
    }
    \label{table:benchmarks}
\end{table}

\begin{table}
  \begin{center}
    \input{sections/table-size-comparison}
  \end{center}
  \caption{The size of each generated concrete configuration normalized to the size of the symbolic configuration it was instantiated from.
    %Except for the bit extraction example, the symbolic configuration was smaller than any of the concrete configurations.
}
  \label{table:size-benchmarks}
\end{table}

%% file: sections/table_program_generation.tex
\begin{tabular*}{\textwidth}{l@{\extracolsep{\fill}}ccccccc}
    \toprule
     &  & \multicolumn{3}{c}{1D (\# PEs)} & \multicolumn{3}{c}{2D ($R\times C$)} \\
    \cmidrule{3-5}\cmidrule{6-8}
    Algorithm & $\pi$ & $1$ & $16$ & $32$ & $4\times 4$ & $8\times 8$ & $32\times 32$ \\
    \midrule
    Bit extract (1D) & 2 & 1.00 (1) & 0.94 (1) & 1.04 (1) & -- & -- & -- \\
    FIR filter (2D) & 2 & 1.00 (1) & 2.64 (3) & 2.72 (3) & 9.06 (9) & 8.98 (9) & 8.23 (9) \\
    Matrix multiplication (3D) & 2 & 1.00 (1) & 0.94 (1) & 0.99 (1) & 1.86 (2) & 1.74 (2) & 1.78 (2) \\
    CNN Convolution (6D) & 1 & 1.00 (1) & 1.77 (2) & 1.78 (2) & 9.22 (4) & 9.33 (4) & 9.06 (4) \\
    \bottomrule
    \end{tabular*}

%% file: sections/table-size-comparison.tex
\begin{tabular*}{\textwidth}{l@{\extracolsep{\fill}}ccccccc}
    \toprule
     &  & \multicolumn{3}{c}{1D (\# PEs)} & \multicolumn{3}{c}{2D ($R\times C$)} \\
    \cmidrule{3-5}\cmidrule{6-8}
    Algorithm & $\pi$ & $1$ & $16$ & $32$ & $4\times 4$ & $8\times 8$ & $32\times 32$ \\
    \midrule
    Bit extract (1D) & 2 & 0.77 & 4.46 & 8.20 & -- & -- & -- \\
    FIR filter (2D) & 2 & 1.17 & 3.86 & 5.85 & 5.14 & 7.18 & 21.36 \\
    Matrix multiplication (3D) & 2 & 1.00 & 3.65 & 6.48 & 4.88 & 15.12 & 193.73 \\
    CNN Convolution (6D) & 1 & 1.05 & 3.73 & 6.53 & 5.83 & 16.25 & 299.41 \\
    \bottomrule
    \end{tabular*}

%% file: sections/conclusion.tex
\section{Conclusion}
\label{sec:conclusion}

In this article, we presented \keyword{symbolic loop compilation}, a two-phase approach that decouples the solving of the NP-complete mapping problem from the actual generation of configuration data, which depends on parameters only known at runtime (the loop bounds and number of available \acp{PE}).
The first phase, \keyword{symbolic mapping}, generates a \keyword{symbolic configuration} that represents the set of concrete configurations over all combinations of loop bounds and numbers \acp{PE}.
The second phase, \keyword{instantiation}, generates a concrete configuration from the symbolic configuration according to the concrete values of the parameters once they become known.

We show that this is a viable approach for dynamically generating configurations because not only does instantiation run in polynomial time, but a symbolic configuration is a very space-efficient representation.
In particular, \keyword{program instantiation}, the most complex part of the instantiation phase, does not directly depend on the number of \acp{PE}, thus scaling to arbitrary sizes of \acp{TCPA}.